\PassOptionsToPackage{unicode}{hyperref}
\PassOptionsToPackage{hyphens}{url}
\PassOptionsToPackage{dvipsnames,svgnames,x11names}{xcolor}
\documentclass[
]{article}
\usepackage{xcolor}
\usepackage{amsmath,amssymb}
\usepackage{iftex}
\ifPDFTeX
  \usepackage[T1]{fontenc}
  \usepackage[utf8]{inputenc}
  \usepackage{textcomp} % provide euro and other symbols
\else % if luatex or xetex
  \usepackage{unicode-math} % this also loads fontspec
  \defaultfontfeatures{Scale=MatchLowercase}
  \defaultfontfeatures[\rmfamily]{Ligatures=TeX,Scale=1}
\fi
\usepackage{lmodern}
\ifPDFTeX\else
\fi
\IfFileExists{upquote.sty}{\usepackage{upquote}}{}
\IfFileExists{microtype.sty}{% use microtype if available
  \usepackage[]{microtype}
  \UseMicrotypeSet[protrusion]{basicmath} % disable protrusion for tt fonts
}{}
\makeatletter
\@ifundefined{KOMAClassName}{% if non-KOMA class
  \IfFileExists{parskip.sty}{%
    \usepackage{parskip}
  }{% else
    \setlength{\parindent}{0pt}
    \setlength{\parskip}{6pt plus 2pt minus 1pt}}
}{% if KOMA class
  \KOMAoptions{parskip=half}}
\makeatother
\makeatletter
\ifx\paragraph\undefined\else
  \let\oldparagraph\paragraph
  \renewcommand{\paragraph}{
    \@ifstar
      \xxxParagraphStar
      \xxxParagraphNoStar
  }
  \newcommand{\xxxParagraphStar}[1]{\oldparagraph*{#1}\mbox{}}
  \newcommand{\xxxParagraphNoStar}[1]{\oldparagraph{#1}\mbox{}}
\fi
\ifx\subparagraph\undefined\else
  \let\oldsubparagraph\subparagraph
  \renewcommand{\subparagraph}{
    \@ifstar
      \xxxSubParagraphStar
      \xxxSubParagraphNoStar
  }
  \newcommand{\xxxSubParagraphStar}[1]{\oldsubparagraph*{#1}\mbox{}}
  \newcommand{\xxxSubParagraphNoStar}[1]{\oldsubparagraph{#1}\mbox{}}
\fi
\makeatother

\usepackage{color}
\usepackage{fancyvrb}

\DefineVerbatimEnvironment{Highlighting}{Verbatim}{commandchars=\\\{\}}
\usepackage{framed}
\definecolor{shadecolor}{RGB}{241,243,245}
\newenvironment{Shaded}{\begin{snugshade}}{\end{snugshade}}

\newcommand{\AttributeTok}[1]{\textcolor[rgb]{0.40,0.45,0.13}{#1}}

\newcommand{\BuiltInTok}[1]{\textcolor[rgb]{0.00,0.23,0.31}{#1}}

\newcommand{\CommentTok}[1]{\textcolor[rgb]{0.37,0.37,0.37}{#1}}

\newcommand{\ConstantTok}[1]{\textcolor[rgb]{0.56,0.35,0.01}{#1}}
\newcommand{\ControlFlowTok}[1]{\textcolor[rgb]{0.00,0.23,0.31}{\textbf{#1}}}
\newcommand{\DataTypeTok}[1]{\textcolor[rgb]{0.68,0.00,0.00}{#1}}

\newcommand{\FloatTok}[1]{\textcolor[rgb]{0.68,0.00,0.00}{#1}}
\newcommand{\FunctionTok}[1]{\textcolor[rgb]{0.28,0.35,0.67}{#1}}
\newcommand{\ImportTok}[1]{\textcolor[rgb]{0.00,0.46,0.62}{#1}}

\newcommand{\KeywordTok}[1]{\textcolor[rgb]{0.00,0.23,0.31}{\textbf{#1}}}
\newcommand{\NormalTok}[1]{\textcolor[rgb]{0.00,0.23,0.31}{#1}}
\newcommand{\OperatorTok}[1]{\textcolor[rgb]{0.37,0.37,0.37}{#1}}
\newcommand{\OtherTok}[1]{\textcolor[rgb]{0.00,0.23,0.31}{#1}}
\newcommand{\PreprocessorTok}[1]{\textcolor[rgb]{0.68,0.00,0.00}{#1}}

\newcommand{\SpecialCharTok}[1]{\textcolor[rgb]{0.37,0.37,0.37}{#1}}

\newcommand{\StringTok}[1]{\textcolor[rgb]{0.13,0.47,0.30}{#1}}

\usepackage{longtable,booktabs,array}
\usepackage{calc} % for calculating minipage widths
\usepackage{etoolbox}
\makeatletter
\patchcmd\longtable{\par}{\if@noskipsec\mbox{}\fi\par}{}{}
\makeatother
\IfFileExists{footnotehyper.sty}{\usepackage{footnotehyper}}{\usepackage{footnote}}
\makesavenoteenv{longtable}
\usepackage{graphicx}
\makeatletter
\newsavebox\pandoc@box
\newcommand*\pandocbounded[1]{% scales image to fit in text height/width
  \sbox\pandoc@box{#1}%
  \Gscale@div\@tempa{\textheight}{\dimexpr\ht\pandoc@box+\dp\pandoc@box\relax}%
  \Gscale@div\@tempb{\linewidth}{\wd\pandoc@box}%
  \ifdim\@tempb\p@<\@tempa\p@\let\@tempa\@tempb\fi% select the smaller of both
  \ifdim\@tempa\p@<\p@\scalebox{\@tempa}{\usebox\pandoc@box}%
  \else\usebox{\pandoc@box}%
  \fi%
}
\def\fps@figure{htbp}
\makeatother

\NewDocumentCommand\citeproctext{}{}

\makeatletter
 \let\@cite@ofmt\@firstofone
 \def\@biblabel#1{}
 \def\@cite#1#2{{#1\if@tempswa , #2\fi}}
\makeatother
\newlength{\cslhangindent}
\newlength{\csllabelwidth}
\newenvironment{CSLReferences}[2] % #1 hanging-indent, #2 entry-spacing
 {\begin{list}{}{%
  \setlength{\itemindent}{0pt}
  \setlength{\leftmargin}{0pt}
  \setlength{\parsep}{0pt}
  \ifodd #1
   \setlength{\leftmargin}{\cslhangindent}
   \setlength{\itemindent}{-1\cslhangindent}
  \fi
  \setlength{\itemsep}{#2\baselineskip}}}
 {\end{list}}
\usepackage{calc}

\providecommand{\tightlist}{%
  \setlength{\itemsep}{0pt}\setlength{\parskip}{0pt}}

\usepackage{arxiv}
\usepackage{orcidlink}
\usepackage{amsmath}
\usepackage[T1]{fontenc}
\makeatletter
\@ifpackageloaded{caption}{}{\usepackage{caption}}
\AtBeginDocument{%
\ifdefined\contentsname
  \renewcommand*\contentsname{Table of contents}
\else
  \newcommand\contentsname{Table of contents}
\fi
\ifdefined\listfigurename
  \renewcommand*\listfigurename{List of Figures}
\else
  \newcommand\listfigurename{List of Figures}
\fi
\ifdefined\listtablename
  \renewcommand*\listtablename{List of Tables}
\else
  \newcommand\listtablename{List of Tables}
\fi
\ifdefined\figurename
  \renewcommand*\figurename{Figure}
\else
  \newcommand\figurename{Figure}
\fi
\ifdefined\tablename
  \renewcommand*\tablename{Table}
\else
  \newcommand\tablename{Table}
\fi
}
\@ifpackageloaded{float}{}{\usepackage{float}}
\floatstyle{ruled}
\@ifundefined{c@chapter}{\newfloat{codelisting}{h}{lop}}{\newfloat{codelisting}{h}{lop}[chapter]}
\floatname{codelisting}{Listing}

\makeatother
\makeatletter
\@ifpackageloaded{caption}{}{\usepackage{caption}}
\@ifpackageloaded{subcaption}{}{\usepackage{subcaption}}
\makeatother
\usepackage{bookmark}
\IfFileExists{xurl.sty}{\usepackage{xurl}}{} % add URL line breaks if available
\hypersetup{
  pdftitle={FormulaCompiler.jl and Margins.jl: Efficient Marginal Effects in Julia},
  pdfauthor={Eric Feltham},
  pdfkeywords={marginal effects, Julia, statistical computing, automatic
differentiation, formula compilation, marginal effects at the
mean, average marginal effects},
  colorlinks=true,
  linkcolor={blue},
  filecolor={Maroon},
  citecolor={Blue},
  urlcolor={Blue},
  pdfcreator={LaTeX via pandoc}}

\newcommand{\runninghead}{A Preprint }
\renewcommand{\runninghead}{Marginal Effects in Julia }
\title{FormulaCompiler.jl and Margins.jl: Efficient Marginal Effects in
Julia}
\author{\textbf{Eric Feltham}\\Data Science Institute and Department of
Sociology\\Columbia
University\\\\\href{mailto:eric.feltham@aya.yale.edu}{eric.feltham@aya.yale.edu}}
\date{}
\begin{document}
\maketitle
\begin{abstract}
Marginal effects analysis is fundamental to interpreting statistical
models, yet existing implementations face computational constraints that
limit analysis at scale. We introduce two Julia packages that address
this gap. Margins.jl provides a clean two-function API organizing
analysis around a \(2 \times 2\) framework: evaluation context
(population vs profile) by analytical target (effects vs predictions).
The package supports interaction analysis through second differences,
elasticity measures, categorical mixtures for representative profiles,
and robust standard errors. FormulaCompiler.jl provides the
computational foundation, transforming statistical formulas into
zero-allocation, type-specialized evaluators that enable \(O(p)\)
per-row computation independent of dataset size. Together, these
packages achieve 622x average speedup and 460x memory reduction compared
to R's marginaleffects package, with successful computation of average
marginal effects and delta-method standard errors on 500,000
observations where R fails due to memory exhaustion---providing the
first comprehensive and efficient marginal effects implementation for
Julia's statistical ecosystem.
\end{abstract}
{\bfseries \emph Keywords}
\def\sep{\textbullet\ }
marginal effects \sep Julia \sep statistical computing \sep automatic
differentiation \sep formula compilation \sep marginal effects at the
mean \sep 
average marginal effects

\section{Introduction}\label{introduction}

Researchers across the social sciences, economics, and biostatistics
increasingly use marginal effects for statistical models, moving beyond
the interpretation of coefficients to quantities that directly answer
substantive questions (Williams 2012; Norton and Dowd 2018; McCabe et
al. 2022; Arel-Bundock, Greifer, and Heiss 2024).

{Julia}'s (Bezanson et al. 2017) statistical ecosystem provides a
framework for statistical formulas ({StatsModels.jl}) and packages for
fitting generalized linear models ({GLM.jl}) and mixed-effects models
({MixedModels.jl}). The {MixedModels.jl} package implements the same
computational approach described in Bates et al. (2015) for the {lme4}
package in {R}, and independent benchmarks report speedups of
approximately 60-fold for certain model specifications (Markwick 2022).
However, no comprehensive package previously existed for computing
marginal effects as a complement to these packages.

Furthermore, existing implementations in {R} face computational
constraints that limit analysis at scale. {R}'s {margins} package
(Leeper {[}2014{]} 2025) encounters performance limitations with large
datasets, with variance estimation identified as the most
computationally expensive operation. While {R}'s newer {marginaleffects}
package (Arel-Bundock, Greifer, and Heiss 2024) achieves substantial
performance improvements---up to 1000x faster than {margins} for some
operations---it remains constrained by \(O(n)\) memory scaling:
computing standard errors via the delta method requires a Jacobian
matrix with \(n\) rows (one per observation)\footnote{\textbf{{Margins.jl}}
  avoids this by accumulating gradients row-by-row into an \(O(p)\)
  buffer instead of materializing the full \(n \times p\) Jacobian.},
and {R}'s underlying \texttt{predict()} methods typically materialize
dense design matrices through \texttt{model.matrix()} (Chambers 1992).
{Python}'s {statsmodels} (Seabold and Perktold 2010) faces analogous
constraints: \texttt{get\_margeff()} creates multiple intermediate
arrays whose sizes scale with \(n\), leading to memory exhaustion when
computing average marginal effects on large datasets. A Python port of
{marginaleffects} exists (Arel-Bundock, Greifer, and Heiss 2024), but
inherits the same scaling characteristics. These constraints can lead to
methodological compromises: computing marginal effects at the mean (MEM)
over average marginal effects (AME) (Williams 2012), subsampling large
datasets, or omitting standard error computation.

\begin{figure}

\centering{

\pandocbounded{\includegraphics[keepaspectratio]{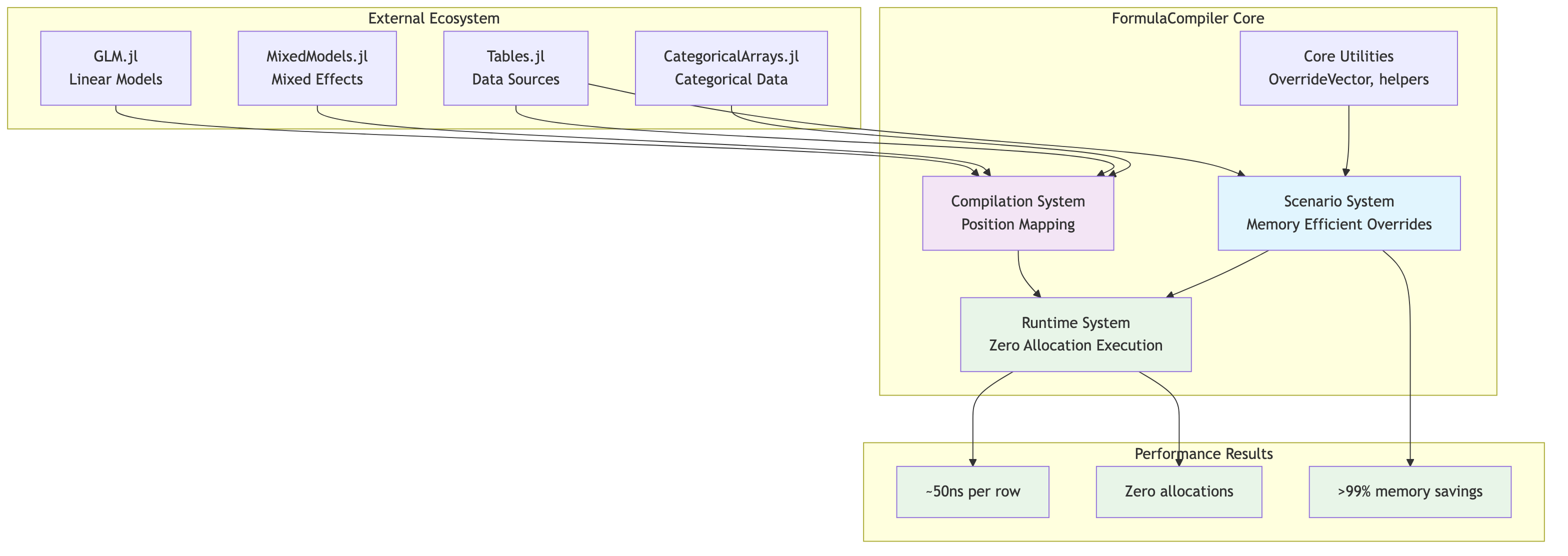}}

}

\caption{\label{fig-ecosystem}System architecture showing {Margins.jl}
providing the statistical interface over {FormulaCompiler.jl}'s
computational engine, with integration to {Julia}'s statistical
ecosystem including {GLM.jl}, {MixedModels.jl}, and {StatsModels.jl}.}

\end{figure}

This paper presents two contributions: (1) a comprehensive marginal
effects package for {Julia}, and (2) an architectural approach that
addresses the memory constraints of existing implementations. Two
complementary packages implement these contributions.
\textbf{{Margins.jl}} provides a statistical interface for marginal
effects analysis---including interactions and categorical
contrasts---through a two-function API that separates
population-averaged analysis (\texttt{population\_margins()}) from
profile-specific analysis (\texttt{profile\_margins()}). This design
makes explicit a distinction that can be obscured in single-function
interfaces: evaluating effects averaged across the observed population
distribution versus evaluating at representative covariate profiles.
\textbf{{FormulaCompiler.jl}} provides the computational foundation
through position-mapped compilation that transforms
{StatsModels.jl}-compatible formulas into zero-allocation,
type-specialized evaluators. The packages integrate with {Julia}'s
statistical ecosystem ({GLM.jl}, {MixedModels.jl}, {StatsModels.jl})
while maintaining the familiar formula interface.
Figure~\ref{fig-ecosystem} illustrates this two-layer architecture.
Benchmarks comparing \textbf{{Margins.jl}} against {R}'s
{marginaleffects} package show substantial improvements in both speed
and memory usage, detailed in the Performance Benchmarks section.

The remainder of this paper proceeds as follows. We first present the
statistical framework for marginal effects analysis that motivates the
\textbf{{Margins.jl}} design. We then describe the \textbf{{Margins.jl}}
API and its \(2 \times 2\) organizational structure. The following
section details how \textbf{{FormulaCompiler.jl}} enables the
performance characteristics through position-mapped compilation and
zero-allocation evaluation. We present benchmarks comparing performance
against {R}'s {marginaleffects} package, followed by case studies
illustrating typical analytical workflows. We conclude with future
directions and a summary of the contributions.

\section{Related Work}\label{related-work}

\subsection{Marginal Effects Software}\label{marginal-effects-software}

The computation of marginal effects has a rich software history across
statistical environments. {Stata}'s \texttt{margins} command (Williams
2012) established the methodological standard, providing comprehensive
functionality for computing average marginal effects (AME), marginal
effects at the mean (MEM), and adjusted predictions within {Stata}'s
integrated environment.

In {R}, the {margins} package (Leeper {[}2014{]} 2025) ported {Stata}'s
functionality to {R} users, establishing API conventions that influenced
subsequent development. The {emmeans} package (Lenth et al. 2025)
introduced reference grid frameworks for factorial designs and estimated
marginal means. Most recently, {marginaleffects} (Arel-Bundock, Greifer,
and Heiss 2024) achieved substantial performance improvements (up to
1000x faster than {margins}). However, even this state-of-the-art
implementation remains constrained by \(O(n)\) memory scaling: the delta
method Jacobian grows with dataset size, and {R}'s \texttt{predict()}
methods materialize design matrices internally.

{Python}'s {statsmodels} (Seabold and Perktold 2010) provides marginal
effects computation through model-specific \texttt{get\_margeff()}
methods, though support is limited to discrete choice models and
standard error computation is not available for all evaluation modes.

In {Julia}, {Effects.jl} (Alday and Kleinschmidt 2024) provides
reference grid predictions and estimated marginal means, but is limited
to profile-based predictions---it does not compute marginal effects
(derivatives) or support population-averaged quantities.

\subsection{Formula Systems}\label{formula-systems}

Statistical formula interfaces based on Wilkinson notation (Wilkinson
and Rogers 1973) represent the standard approach for model
specification. {R}'s \texttt{model.matrix()} (Chambers and Hastie 1992),
{Python}'s {Patsy} (Smith 2011), and {Julia}'s {StatsModels.jl}
(Kleinschmidt et al. 2024) all implement variants of this framework.
These implementations follow a common pattern: parse the formula
specification, then materialize a full design matrix \(O(n \times p)\)
for the entire dataset. This approach proves effective for model
estimation but imposes substantial overhead for workloads that involve
repeated per-row evaluation, including marginal effects computation.

\subsection{Positioning}\label{positioning}

\textbf{{Margins.jl}} and \textbf{{FormulaCompiler.jl}} address this gap
between statistical usability and computational performance. Unlike
existing marginal effects packages, \textbf{{Margins.jl}} explicitly
organizes analysis around the \(2 \times 2\) framework of evaluation
context and analytical target, providing clarity about what quantities
are being computed. Unlike existing formula systems,
\textbf{{FormulaCompiler.jl}} eliminates design matrix materialization
through position-mapped compilation, enabling \(O(p)\) per-row
evaluation. Together, these packages enable analysis at scales difficult
or impossible with existing implementations, while maintaining a
familiar interface.

\section{Marginal Effects: Statistical
Framework}\label{marginal-effects-statistical-framework}

This section establishes the statistical foundation for marginal effects
analysis, introducing the \(2 \times 2\) framework that organizes the
\textbf{{Margins.jl}} design.

We consider models expressed with {StatsModels.jl} formulas. Let \(p\)
denote the number of fixed-effect terms after contrast expansion.

\begin{itemize}
\tightlist
\item
  Linear models: \(E[y_i] = x_i^T \beta\), where
  \(x_i \in \mathbb{R}^p\)
\item
  Generalized linear models (GLMs): \(\mu_i = g^{-1}(\eta_i)\),
  \(\eta_i = x_i^T \beta\)
\item
  Mixed models: we target fixed effects only, extracting \(x_i^T \beta\)
  from {MixedModels.jl} fits
\end{itemize}

Categorical predictors are expanded according to the contrast coding
selected. Interactions, transformations, and boolean terms follow
standard formula semantics.

Marginal effects quantify the expected change in an outcome resulting
from a unit change in an explanatory variable. For continuous
covariates, this is formally defined as
\(\frac{\partial E[Y|X]}{\partial x_k}\), the partial derivative of the
conditional expectation with respect to covariate \(x_k\). For
categorical variables, marginal effects are computed as contrasts
between levels.

For GLMs with link function \(g\), the marginal effect on the response
scale involves the chain rule:

\[\frac{\partial \mu}{\partial x_k} = \frac{d\mu}{d\eta} \cdot \frac{\partial \eta}{\partial x_k} = g'^{-1}(\eta) \cdot \frac{\partial (x^T\beta)}{\partial x_k}\]

where \((g^{-1})'(\eta) = \frac{d\mu}{d\eta}\) is the derivative of the
inverse link function.

Adjusted predictions represent the conditional expectation \(E[Y|X]\)
evaluated at specified covariate configurations, providing fitted values
at particular conditions.

\subsection{\texorpdfstring{The \(2 \times 2\)
Framework}{The 2 \textbackslash times 2 Framework}}\label{the-2-times-2-framework}

The marginal effects analysis rests upon two axes that define the
distinct quantities that researchers may choose to estimate
(Table~\ref{tbl-framework}):

\begin{longtable}[]{@{}
  >{\raggedright\arraybackslash}p{(\linewidth - 4\tabcolsep) * \real{0.2222}}
  >{\raggedright\arraybackslash}p{(\linewidth - 4\tabcolsep) * \real{0.3556}}
  >{\raggedright\arraybackslash}p{(\linewidth - 4\tabcolsep) * \real{0.4222}}@{}}
\caption{The \(2 \times 2\) framework for marginal effects analysis:
evaluation context (rows) crossed with analytical target (columns).
Profile analysis encompasses two variants: evaluation at sample means
(MEM/APM) and evaluation at arbitrary reference profiles
(MER/APR).}\label{tbl-framework}\tabularnewline
\toprule\noalign{}
\begin{minipage}[b]{\linewidth}\raggedright
\textbf{Context} \textbackslash{} \textbf{Target}
\end{minipage} & \begin{minipage}[b]{\linewidth}\raggedright
\textbf{Effects}
\end{minipage} & \begin{minipage}[b]{\linewidth}\raggedright
\textbf{Predictions}
\end{minipage} \\
\midrule\noalign{}
\endfirsthead
\toprule\noalign{}
\begin{minipage}[b]{\linewidth}\raggedright
\textbf{Context} \textbackslash{} \textbf{Target}
\end{minipage} & \begin{minipage}[b]{\linewidth}\raggedright
\textbf{Effects}
\end{minipage} & \begin{minipage}[b]{\linewidth}\raggedright
\textbf{Predictions}
\end{minipage} \\
\midrule\noalign{}
\endhead
\bottomrule\noalign{}
\endlastfoot
\textbf{Population} & Average Marginal Effects (AME) & Average Adjusted
Predictions (AAP) \\
\textbf{Profile (at means)} & Marginal Effects at the Mean (MEM) &
Adjusted Predictions at the Mean (APM) \\
\textbf{Profile (at reference)} & Marginal Effects at Reference (MER) &
Adjusted Predictions at Reference (APR) \\
\end{longtable}

Evaluation context determines whether inference targets population-level
parameters or profile-specific estimates:

\begin{itemize}
\tightlist
\item
  Population approaches compute marginal quantities averaged across the
  empirical distribution of observed covariates, yielding
  population-representative estimates that reflect heterogeneity in the
  data.
\item
  Profile approaches evaluate marginal quantities at specific covariate
  configurations defined by a reference grid. The \texttt{means\_grid()}
  helper produces evaluation at sample means (MEM/APM), representing a
  ``typical'' observation. Custom grids (see Section~\ref{sec-profile})
  via \texttt{cartesian\_grid()} or user-supplied DataFrames produce
  evaluation at arbitrary reference profiles (MER/APR), representing
  researcher-specified counterfactual scenarios. Both use
  \texttt{profile\_margins()} but differ in interpretation.
\end{itemize}

Analytical target specifies whether the quantity of interest represents
effects (derivatives/contrasts) or predictions (fitted values).

For linear models without interactions, marginal effects are constant
(equal to coefficients), so profile and population approaches yield
identical results. With interactions, AME equals MEM (effects at
covariate means) because marginal effects remain linear in other
covariates, but effects at arbitrary profiles will differ. For nonlinear
models (logit, probit, etc.), the distinction becomes substantively
important regardless of model specification:

\begin{itemize}
\tightlist
\item
  AME provides population-representative effect estimates by averaging
  over observed covariate combinations
\item
  MEM provides effect estimates at a specific (potentially
  unrepresentative) point
\end{itemize}

Researchers increasingly favor AME over MEM for this reason
(Arel-Bundock, Greifer, and Heiss 2024; Hanmer and Kalkan 2013), but
AME's statistical superiority comes at computational cost: \(O(n)\)
marginal effect calculations compared to \(O(1)\) for MEM. The
performance limitations of existing software have historically pushed
researchers toward MEM as an optimization strategy, illustrating how
computational constraints influence methodological choices.

\subsection{Variance Estimation}\label{variance-estimation}

Standard errors for marginal effects are computed via the delta method.
For a scalar marginal effect \(m(\beta)\):

\[\text{Var}(m(\hat{\beta})) \approx g^T \Sigma g\]

where \(g = \frac{\partial m}{\partial \beta}\) is the gradient and
\(\Sigma\) is the parameter covariance matrix.

For AME, this requires accumulating gradients across observations:

\[g_{\text{AME}} = \frac{1}{n} \sum_i g_i\]

Efficient computation of these quantities---particularly the
per-observation gradients---is central to enabling statistical inference
at scale, and motivates the computational architecture described in
subsequent sections.

\section{Margins.jl: Statistical
Interface}\label{margins.jl-statistical-interface}

\textbf{{Margins.jl}} translates the \(2 \times 2\) statistical
framework into a clean two-function API that makes the analytical
choices explicit. This section describes the API design, key features,
and typical workflows.

\subsection{API Design: Two Functions}\label{api-design-two-functions}

\textbf{{Margins.jl}} separates analysis into two entry points
corresponding to the evaluation context, differing from other packages
that provide may one function with many options (as in {R}'s {margins}
or {marginaleffects}),

\begin{Shaded}
\begin{Highlighting}[]
\CommentTok{\# Population analysis: average across observed sample distribution}
\FunctionTok{population\_margins}\NormalTok{(model, data; }\KeywordTok{type}\OperatorTok{=:}\NormalTok{effects)  }\CommentTok{\# AME}
\FunctionTok{population\_margins}\NormalTok{(model, data; }\KeywordTok{type}\OperatorTok{=:}\NormalTok{predictions)  }\CommentTok{\# AAP}

\CommentTok{\# Profile analysis: evaluate at specific covariate configurations}
\FunctionTok{profile\_margins}\NormalTok{(model, data, reference\_grid; }\KeywordTok{type}\OperatorTok{=:}\NormalTok{effects)  }\CommentTok{\# MEM}
\FunctionTok{profile\_margins}\NormalTok{(model, data, reference\_grid; }\KeywordTok{type}\OperatorTok{=:}\NormalTok{predictions)  }\CommentTok{\# APM}
\end{Highlighting}
\end{Shaded}

For profile analysis, \texttt{profile\_margins()} computes effects or
predictions at specific covariate configurations defined by a reference
grid. When \texttt{type=:effects}, the function evaluates the marginal
effect (derivative for continuous variables, contrast for categorical
variables) at each grid point. Unlike population analysis, which
averages effects across all observed covariate combinations, profile
analysis evaluates at the exact covariate values specified in the
grid---producing marginal effects at the mean (MEM) when the grid
contains sample means, or marginal effects at arbitrary reference points
(MER) for custom grids.

For categorical variables, profile analysis computes contrasts at each
grid point (\emph{e.g.}, ``B'' vs ``A'') instead of averaging contrasts
across the sample. Binary variables are treated as categorical, with
effects computed as the discrete change in predicted outcome when
switching levels.

This separation provides several benefits:

\begin{enumerate}
\def\labelenumi{\arabic{enumi}.}
\tightlist
\item
  Clarity: Users explicitly choose between population and profile
  analysis
\item
  Appropriate defaults: Profile analysis requires a reference grid;
  population analysis does not
\item
  Documentation: Each function documents only its relevant parameters
\item
  Performance: Implementation can optimize for each case separately
\end{enumerate}

\subsection{Population Analysis}\label{population-analysis}

The \texttt{population\_margins()} function computes quantities averaged
across the observed sample distribution:

\begin{Shaded}
\begin{Highlighting}[]
\ImportTok{using} \BuiltInTok{Margins}\NormalTok{, }\BuiltInTok{GLM}\NormalTok{, }\BuiltInTok{DataFrames}

\CommentTok{\# Fit model}
\NormalTok{model }\OperatorTok{=} \FunctionTok{glm}\NormalTok{(}\PreprocessorTok{@formula}\NormalTok{(y }\OperatorTok{\textasciitilde{}}\NormalTok{ x1 }\OperatorTok{+}\NormalTok{ x2 }\OperatorTok{+}\NormalTok{ group), data, }\FunctionTok{Binomial}\NormalTok{())}

\CommentTok{\# Average marginal effects for all continuous variables}
\NormalTok{ame\_result }\OperatorTok{=} \FunctionTok{population\_margins}\NormalTok{(model, data; }\KeywordTok{type}\OperatorTok{=:}\NormalTok{effects)}
\end{Highlighting}
\end{Shaded}

\begin{Shaded}
\begin{Highlighting}[]
\NormalTok{EffectsResult}\OperatorTok{:} \FloatTok{4}\NormalTok{ population effects (N}\OperatorTok{=}\FloatTok{5000}\NormalTok{)}
\OperatorTok{{-}{-}{-}{-}{-}{-}{-}{-}{-}{-}{-}{-}{-}{-}{-}{-}{-}{-}{-}{-}{-}{-}{-}{-}{-}{-}{-}{-}{-}{-}{-}{-}{-}{-}{-}{-}{-}{-}{-}{-}{-}{-}{-}{-}{-}{-}{-}{-}{-}{-}{-}{-}{-}{-}{-}{-}{-}{-}{-}{-}{-}{-}{-}{-}{-}{-}{-}{-}{-}{-}{-}}
\NormalTok{Variable Contrast         dy}\OperatorTok{/}\NormalTok{dx   Std. Err.  [}\FloatTok{95}\OperatorTok{\%}\NormalTok{ Conf.   Interval]}
\OperatorTok{{-}{-}{-}{-}{-}{-}{-}{-}{-}{-}{-}{-}{-}{-}{-}{-}{-}{-}{-}{-}{-}{-}{-}{-}{-}{-}{-}{-}{-}{-}{-}{-}{-}{-}{-}{-}{-}{-}{-}{-}{-}{-}{-}{-}{-}{-}{-}{-}{-}{-}{-}{-}{-}{-}{-}{-}{-}{-}{-}{-}{-}{-}{-}{-}{-}{-}{-}{-}{-}{-}{-}}
\NormalTok{x1       dy}\OperatorTok{/}\NormalTok{dx         }\OperatorTok{{-}}\FloatTok{0.00737}     \FloatTok{0.00687}    \OperatorTok{{-}}\FloatTok{0.02084}      \FloatTok{0.0061}
\NormalTok{x2       dy}\OperatorTok{/}\NormalTok{dx          }\FloatTok{0.00145}     \FloatTok{0.00685}    \OperatorTok{{-}}\FloatTok{0.01198}     \FloatTok{0.01487}
\NormalTok{group    B }\OperatorTok{{-}}\NormalTok{ A         }\OperatorTok{{-}}\FloatTok{0.02436}     \FloatTok{0.01686}    \OperatorTok{{-}}\FloatTok{0.05741}     \FloatTok{0.00869}
\NormalTok{group    C }\OperatorTok{{-}}\NormalTok{ A          }\FloatTok{0.01283}     \FloatTok{0.01683}    \OperatorTok{{-}}\FloatTok{0.02015}     \FloatTok{0.04581}
\OperatorTok{{-}{-}{-}{-}{-}{-}{-}{-}{-}{-}{-}{-}{-}{-}{-}{-}{-}{-}{-}{-}{-}{-}{-}{-}{-}{-}{-}{-}{-}{-}{-}{-}{-}{-}{-}{-}{-}{-}{-}{-}{-}{-}{-}{-}{-}{-}{-}{-}{-}{-}{-}{-}{-}{-}{-}{-}{-}{-}{-}{-}{-}{-}{-}{-}{-}{-}{-}{-}{-}{-}{-}}
\end{Highlighting}
\end{Shaded}

\begin{Shaded}
\begin{Highlighting}[]
\CommentTok{\# AME for specific variables only}
\NormalTok{ame\_x1 }\OperatorTok{=} \FunctionTok{population\_margins}\NormalTok{(model, data; }\KeywordTok{type}\OperatorTok{=:}\NormalTok{effects, vars}\OperatorTok{=}\NormalTok{[}\OperatorTok{:}\NormalTok{x1])}

\CommentTok{\# AME at counterfactual scenario values}
\NormalTok{ame\_scenario }\OperatorTok{=} \FunctionTok{population\_margins}\NormalTok{(}
\NormalTok{    model, data; }\KeywordTok{type}\OperatorTok{=:}\NormalTok{effects, vars}\OperatorTok{=}\NormalTok{[}\OperatorTok{:}\NormalTok{x1], scenarios}\OperatorTok{=}\NormalTok{(x2}\OperatorTok{=}\FloatTok{1.0}\NormalTok{,)}
\NormalTok{)}

\CommentTok{\# AME stratified by groups}
\NormalTok{ame\_grouped }\OperatorTok{=} \FunctionTok{population\_margins}\NormalTok{(model, data; }\KeywordTok{type}\OperatorTok{=:}\NormalTok{effects, vars}\OperatorTok{=}\NormalTok{[}\OperatorTok{:}\NormalTok{x1], groups}\OperatorTok{=:}\NormalTok{region)}
\end{Highlighting}
\end{Shaded}

The \texttt{scenarios} parameter enables counterfactual analysis by
overriding specific variable values while computing effects across the
sample distribution. The \texttt{groups} parameter enables subgroup
analysis, computing separate AME estimates for each level of a grouping
variable. Weighted analysis is supported via the \texttt{weights}
parameter, which incorporates observation weights into both the marginal
effect computation and variance estimation. For survey data, weights
adjust the AME to represent the target population rather than the sample
(Winship and Radbill 1994; Lumley 2004). The weighted AME is computed as
\(\bar{\partial} = \sum_i w_i \cdot \partial_i / \sum_i w_i\), where
\(w_i\) are the observation weights.

\subsection{Profile Analysis and Reference Grids}\label{sec-profile}

The \texttt{profile\_margins()} function evaluates at specific covariate
configurations defined by a reference grid:

\begin{Shaded}
\begin{Highlighting}[]
\CommentTok{\# Effects at sample means}
\NormalTok{mem\_result }\OperatorTok{=} \FunctionTok{profile\_margins}\NormalTok{(model, data, }\FunctionTok{means\_grid}\NormalTok{(data); }\KeywordTok{type}\OperatorTok{=:}\NormalTok{effects)}

\CommentTok{\# Effects at custom grid points}
\NormalTok{grid }\OperatorTok{=} \FunctionTok{cartesian\_grid}\NormalTok{(}
\NormalTok{    x1 }\OperatorTok{=}\NormalTok{ [}\FloatTok{0.0}\NormalTok{, }\FloatTok{1.0}\NormalTok{, }\FloatTok{2.0}\NormalTok{],}
\NormalTok{    x2 }\OperatorTok{=}\NormalTok{ [}\OperatorTok{{-}}\FloatTok{1.0}\NormalTok{, }\FloatTok{0.0}\NormalTok{, }\FloatTok{1.0}\NormalTok{]}
\NormalTok{)}
\NormalTok{effects\_at\_grid }\OperatorTok{=} \FunctionTok{profile\_margins}\NormalTok{(model, data, grid; }\KeywordTok{type}\OperatorTok{=:}\NormalTok{effects)}

\CommentTok{\# Predictions across a profile sweep}
\NormalTok{pred\_grid }\OperatorTok{=} \FunctionTok{profile\_margins}\NormalTok{(model, data, grid; }\KeywordTok{type}\OperatorTok{=:}\NormalTok{predictions)}
\end{Highlighting}
\end{Shaded}

\textbf{{Margins.jl}} provides several reference grid builders:

\begin{itemize}
\tightlist
\item
  \texttt{means\_grid()}: Sample means for continuous variables
\item
  \texttt{cartesian\_grid()}: Cartesian product of specified values
\item
  \texttt{balanced\_grid()}: Balanced design for categorical variables
\item
  \texttt{quantile\_grid()}: Quantile-based grid points
\end{itemize}

\begin{Shaded}
\begin{Highlighting}[]
\CommentTok{\# Example: Create a grid over two variables}
\NormalTok{grid }\OperatorTok{=} \FunctionTok{cartesian\_grid}\NormalTok{(}
\NormalTok{    income }\OperatorTok{=}\NormalTok{ [}\FloatTok{25000}\NormalTok{, }\FloatTok{50000}\NormalTok{, }\FloatTok{75000}\NormalTok{, }\FloatTok{100000}\NormalTok{],}
\NormalTok{    education }\OperatorTok{=}\NormalTok{ [}\StringTok{"HS"}\NormalTok{, }\StringTok{"BA"}\NormalTok{, }\StringTok{"MA"}\NormalTok{]}
\NormalTok{)}
\CommentTok{\# Returns 12{-}row DataFrame: all combinations of income × education}
\end{Highlighting}
\end{Shaded}

Additional specialized builders are available (\emph{e.g.},
\texttt{hierarchical\_grid()} for nested designs). Users can also supply
any DataFrame directly as a custom reference grid.

\subsection{Categorical Variables}\label{categorical-variables}

\textbf{{Margins.jl}} handles categorical variables automatically
through contrast-based computation. Effects are computed as contrasts,
with \texttt{:baseline} (vs reference level, default) or
\texttt{:pairwise} (all pairwise comparisons) contrast types.
Predictions are computed at each level of the categorical variable.

\begin{Shaded}
\begin{Highlighting}[]
\CommentTok{\# Model with categorical predictor}
\NormalTok{model }\OperatorTok{=} \FunctionTok{lm}\NormalTok{(}\PreprocessorTok{@formula}\NormalTok{(y }\OperatorTok{\textasciitilde{}}\NormalTok{ x }\OperatorTok{+}\NormalTok{ group), data)}

\CommentTok{\# AME automatically detects and handles categorical \textquotesingle{}group\textquotesingle{}}
\NormalTok{result }\OperatorTok{=} \FunctionTok{population\_margins}\NormalTok{(model, data; }\KeywordTok{type}\OperatorTok{=:}\NormalTok{effects)}
\end{Highlighting}
\end{Shaded}

\begin{verbatim}
EffectsResult: 3 population effects (N=5000)
-----------------------------------------------------------------------
Variable Contrast         dy/dx   Std. Err.  [95% Conf.   Interval]
-----------------------------------------------------------------------
x        dy/dx          0.39729     0.00711     0.38335     0.41122
group    B - A          0.37231     0.01751       0.338     0.40662
group    C - A          0.59665     0.01731     0.56272     0.63058
-----------------------------------------------------------------------
\end{verbatim}

The output shows continuous variable \texttt{x} with its derivative
(\texttt{dy/dx}), and categorical variable \texttt{group} with contrasts
versus the baseline level A.

\subsection{Categorical Mixtures}\label{categorical-mixtures}

For profile analysis, \textbf{{Margins.jl}} supports categorical
mixtures---weighted combinations of categorical levels that avoid
arbitrary ``representative value'' choices. By default, categorical
variables use observed proportions as mixture weights. For example, if a
sample contains 60\% group A and 40\% group B, the default mixture
computes effects as
\(0.6 \times \text{effect}(A) + 0.4 \times \text{effect}(B)\). Custom
weights can override observed proportions for sensitivity analysis or to
match external population distributions.

\begin{Shaded}
\begin{Highlighting}[]
\CommentTok{\# Instead of picking a single group level...}
\CommentTok{\# Use frequency{-}weighted mixture of observed proportions}
\NormalTok{grid }\OperatorTok{=} \FunctionTok{means\_grid}\NormalTok{(data; group}\OperatorTok{=:}\NormalTok{mixture)}
\NormalTok{mem\_mixture }\OperatorTok{=} \FunctionTok{profile\_margins}\NormalTok{(model, data, grid; }\KeywordTok{type}\OperatorTok{=:}\NormalTok{effects)}
\end{Highlighting}
\end{Shaded}

Internally, mixtures are represented as weighted level combinations:

\begin{Shaded}
\begin{Highlighting}[]
\KeywordTok{struct}\NormalTok{ CategoricalMixture\{T\}}
\NormalTok{    levels}\OperatorTok{::}\DataTypeTok{Vector\{T\}}
\NormalTok{    weights}\OperatorTok{::}\DataTypeTok{Vector\{Float64\}}
\KeywordTok{end}
\end{Highlighting}
\end{Shaded}

This allows profile-based marginal effects to weight across categorical
levels according to sample frequencies, providing more representative
estimates than selecting a single reference level.

\subsection{Result Types and Output}\label{result-types-and-output}

Both functions return structured result types that organize output for
downstream analysis and presentation:

\begin{Shaded}
\begin{Highlighting}[]
\KeywordTok{struct}\NormalTok{ EffectsResult }\OperatorTok{\textless{}:}\DataTypeTok{ MarginsResult}
\NormalTok{    estimates}\OperatorTok{::}\DataTypeTok{Vector\{Float64\}}
\NormalTok{    standard\_errors}\OperatorTok{::}\DataTypeTok{Vector\{Float64\}}
\NormalTok{    variables}\OperatorTok{::}\DataTypeTok{Vector\{String\}}
\NormalTok{    terms}\OperatorTok{::}\DataTypeTok{Vector\{String\}}
\NormalTok{    profile\_values}\OperatorTok{::}\DataTypeTok{Union\{Nothing, NamedTuple\}}
\NormalTok{    group\_values}\OperatorTok{::}\DataTypeTok{Union\{Nothing, NamedTuple\}}
\NormalTok{    gradients}\OperatorTok{::}\DataTypeTok{Matrix\{Float64\}}
\NormalTok{    metadata}\OperatorTok{::}\DataTypeTok{Dict\{Symbol, Any\}}
\KeywordTok{end}

\KeywordTok{struct}\NormalTok{ PredictionsResult }\OperatorTok{\textless{}:}\DataTypeTok{ MarginsResult}
\NormalTok{    estimates}\OperatorTok{::}\DataTypeTok{Vector\{Float64\}}
\NormalTok{    standard\_errors}\OperatorTok{::}\DataTypeTok{Vector\{Float64\}}
\NormalTok{    profile\_values}\OperatorTok{::}\DataTypeTok{Union\{Nothing, NamedTuple\}}
\NormalTok{    group\_values}\OperatorTok{::}\DataTypeTok{Union\{Nothing, NamedTuple\}}
\NormalTok{    gradients}\OperatorTok{::}\DataTypeTok{Matrix\{Float64\}}
\NormalTok{    metadata}\OperatorTok{::}\DataTypeTok{Dict\{Symbol, Any\}}
\KeywordTok{end}

\KeywordTok{struct}\NormalTok{ ContrastResult}
\NormalTok{    contrast}\OperatorTok{::}\DataTypeTok{Float64}
\NormalTok{    se}\OperatorTok{::}\DataTypeTok{Float64}
\NormalTok{    t\_stat}\OperatorTok{::}\DataTypeTok{Float64}
\NormalTok{    p\_value}\OperatorTok{::}\DataTypeTok{Float64}
\NormalTok{    ci\_lower}\OperatorTok{::}\DataTypeTok{Float64}
\NormalTok{    ci\_upper}\OperatorTok{::}\DataTypeTok{Float64}
\NormalTok{    estimate1}\OperatorTok{::}\DataTypeTok{Float64}
\NormalTok{    estimate2}\OperatorTok{::}\DataTypeTok{Float64}
\NormalTok{    gradient}\OperatorTok{::}\DataTypeTok{Vector\{Float64\}}
\NormalTok{    row1}\OperatorTok{::}\DataTypeTok{Union\{Int, Nothing\}}
\NormalTok{    row2}\OperatorTok{::}\DataTypeTok{Union\{Int, Nothing\}}
\NormalTok{    metadata}\OperatorTok{::}\DataTypeTok{Dict\{Symbol, Any\}}
\KeywordTok{end}
\end{Highlighting}
\end{Shaded}

\texttt{EffectsResult} contains marginal effects (AME, MEM, or MER) with
delta-method standard errors, while \texttt{PredictionsResult} contains
adjusted predictions (AAP, APM, or APR). Both store gradient matrices to
enable subsequent contrast computation, and
\texttt{profile\_values}/\texttt{group\_values} capture the evaluation
context. \texttt{ContrastResult} is returned by the \texttt{contrast()}
function when comparing two estimates (e.g., predictions at different
covariate values or effects across groups), providing complete
statistical inference for the difference. These types convert naturally
to DataFrames:

\begin{Shaded}
\begin{Highlighting}[]
\NormalTok{result }\OperatorTok{=} \FunctionTok{population\_margins}\NormalTok{(model, data; }\KeywordTok{type}\OperatorTok{=:}\NormalTok{effects)}

\CommentTok{\# Access individual components}
\NormalTok{result.estimates         }\CommentTok{\# Point estimates}
\NormalTok{result.standard\_errors   }\CommentTok{\# Standard errors}
\NormalTok{result.variables         }\CommentTok{\# Variable names}

\CommentTok{\# Access as DataFrame}
\NormalTok{df }\OperatorTok{=} \FunctionTok{DataFrame}\NormalTok{(result)}
\end{Highlighting}
\end{Shaded}

\begin{Shaded}
\begin{Highlighting}[]
\FloatTok{4}\NormalTok{x10 DataFrame}
\NormalTok{ Row }\OperatorTok{|} \KeywordTok{type}\NormalTok{    variable  contrast  estimate     se          t\_stat     p\_value   ci\_lower    ci\_upper    n}
     \OperatorTok{|} \DataTypeTok{String}  \DataTypeTok{String}    \DataTypeTok{String}    \DataTypeTok{Float64}      \DataTypeTok{Float64}     \DataTypeTok{Float64}    \DataTypeTok{Float64}   \DataTypeTok{Float64}     \DataTypeTok{Float64}     \DataTypeTok{Int64}
\OperatorTok{{-}{-}{-}{-}{-}+{-}{-}{-}{-}{-}{-}{-}{-}{-}{-}{-}{-}{-}{-}{-}{-}{-}{-}{-}{-}{-}{-}{-}{-}{-}{-}{-}{-}{-}{-}{-}{-}{-}{-}{-}{-}{-}{-}{-}{-}{-}{-}{-}{-}{-}{-}{-}{-}{-}{-}{-}{-}{-}{-}{-}{-}{-}{-}{-}{-}{-}{-}{-}{-}{-}{-}{-}{-}{-}{-}{-}{-}{-}{-}{-}{-}{-}{-}{-}{-}{-}{-}{-}{-}{-}{-}{-}{-}{-}{-}{-}{-}{-}{-}{-}{-}{-}{-}{-}{-}{-}{-}{-}{-}{-}}
   \FloatTok{1} \OperatorTok{|}\NormalTok{ AME     x1        dy}\OperatorTok{/}\NormalTok{dx     }\OperatorTok{{-}}\FloatTok{0.00737132}  \FloatTok{0.0068739}   \OperatorTok{{-}}\FloatTok{1.07236}   \FloatTok{0.283557}  \OperatorTok{{-}}\FloatTok{0.0208439}  \FloatTok{0.00610128}   \FloatTok{5000}
   \FloatTok{2} \OperatorTok{|}\NormalTok{ AME     x2        dy}\OperatorTok{/}\NormalTok{dx      }\FloatTok{0.00144605}  \FloatTok{0.00685147}   \FloatTok{0.211057}  \FloatTok{0.832843}  \OperatorTok{{-}}\FloatTok{0.0119826}  \FloatTok{0.0148747}    \FloatTok{5000}
   \FloatTok{3} \OperatorTok{|}\NormalTok{ AME     group     B }\OperatorTok{{-}}\NormalTok{ A     }\OperatorTok{{-}}\FloatTok{0.0243615}   \FloatTok{0.0168627}   \OperatorTok{{-}}\FloatTok{1.4447}    \FloatTok{0.148542}  \OperatorTok{{-}}\FloatTok{0.0574117}  \FloatTok{0.00868869}   \FloatTok{5000}
   \FloatTok{4} \OperatorTok{|}\NormalTok{ AME     group     C }\OperatorTok{{-}}\NormalTok{ A      }\FloatTok{0.0128309}   \FloatTok{0.0168256}    \FloatTok{0.762581}  \FloatTok{0.445713}  \OperatorTok{{-}}\FloatTok{0.0201467}  \FloatTok{0.0458084}    \FloatTok{5000}
\end{Highlighting}
\end{Shaded}

The DataFrame includes columns for \texttt{type} (AME/MEM),
\texttt{variable}, \texttt{contrast}, point estimates, standard errors,
test statistics, p-values, confidence intervals, and sample size.

The \texttt{contrast()} function computes differences between estimates
with proper delta-method inference:

\begin{Shaded}
\begin{Highlighting}[]
\CommentTok{\# Get profile effects with gradient information}
\NormalTok{result }\OperatorTok{=} \FunctionTok{profile\_margins}\NormalTok{(model, data,}
    \FunctionTok{cartesian\_grid}\NormalTok{(education}\OperatorTok{=}\NormalTok{[}\StringTok{"HS"}\NormalTok{, }\StringTok{"College"}\NormalTok{, }\StringTok{"Grad"}\NormalTok{]);}
    \KeywordTok{type}\OperatorTok{=:}\NormalTok{effects, vars}\OperatorTok{=}\NormalTok{[}\OperatorTok{:}\NormalTok{x1])}
\NormalTok{df }\OperatorTok{=} \FunctionTok{DataFrame}\NormalTok{(result; include\_gradients}\OperatorTok{=}\ConstantTok{true}\NormalTok{)}

\CommentTok{\# Contrast by row indices}
\NormalTok{cr }\OperatorTok{=} \FunctionTok{contrast}\NormalTok{(df, }\FloatTok{2}\NormalTok{, }\FloatTok{1}\NormalTok{, }\FunctionTok{vcov}\NormalTok{(model))  }\CommentTok{\# Row 2 vs Row 1}

\CommentTok{\# Contrast by column specifications (NamedTuple)}
\NormalTok{cr }\OperatorTok{=} \FunctionTok{contrast}\NormalTok{(df,}
\NormalTok{    (variable}\OperatorTok{=:}\NormalTok{x1, education}\OperatorTok{=}\StringTok{"College"}\NormalTok{),}
\NormalTok{    (variable}\OperatorTok{=:}\NormalTok{x1, education}\OperatorTok{=}\StringTok{"HS"}\NormalTok{),}
    \FunctionTok{vcov}\NormalTok{(model))}
\end{Highlighting}
\end{Shaded}

\subsection{Elasticities}\label{elasticities}

Elasticities express marginal effects in percentage terms, and not
absolute units, facilitating comparison across variables with different
scales. The elasticity of \(Y\) with respect to \(X\) measures the
percentage change in \(Y\) for a one percent change in \(X\):
\(\varepsilon = \frac{\partial Y}{\partial X} \cdot \frac{X}{Y}\).
Semi-elasticities (\texttt{eyex} and \texttt{dyex}) provide hybrid
measures where only one side is in percentage terms (Stock and Watson
2011).

\textbf{{Margins.jl}} supports elasticity variants through the
\texttt{measure} parameter:

\begin{Shaded}
\begin{Highlighting}[]
\CommentTok{\# Standard elasticity: (dy/dx) * (x/y)}
\FunctionTok{population\_margins}\NormalTok{(model, data; }\KeywordTok{type}\OperatorTok{=:}\NormalTok{effects, measure}\OperatorTok{=:}\NormalTok{elasticity)}

\CommentTok{\# Semi{-}elasticity dyex: (dy/dx) * x}
\FunctionTok{population\_margins}\NormalTok{(model, data; }\KeywordTok{type}\OperatorTok{=:}\NormalTok{effects, measure}\OperatorTok{=:}\NormalTok{dyex)}

\CommentTok{\# Semi{-}elasticity eydx: (dy/dx) / y}
\FunctionTok{population\_margins}\NormalTok{(model, data; }\KeywordTok{type}\OperatorTok{=:}\NormalTok{effects, measure}\OperatorTok{=:}\NormalTok{eydx)}
\end{Highlighting}
\end{Shaded}

\subsection{Interaction Effects and Second
Differences}\label{interaction-effects-and-second-differences}

When models include interaction terms, researchers often want to know
whether the marginal effect of one variable depends on the level of
another (Ai and Norton 2003; Brambor, Clark, and Golder 2006).
\textbf{Second differences} quantify this dependence: the difference in
marginal effects across levels of a moderating variable. For continuous
moderators, this extends to derivatives of marginal effects with respect
to the moderator.

In linear models, second differences equal the interaction coefficient
directly. In nonlinear models (logit, probit, Poisson), however,
interaction coefficients exist on the link scale and do not represent
interactions on the predicted outcome scale (McCabe et al. 2022). Proper
interpretation requires computing second differences from marginal
effects:

\[\text{Second Difference} = \frac{\partial E[Y|X, Z_{\text{high}}]}{\partial X} - \frac{\partial E[Y|X, Z_{\text{low}}]}{\partial X}\]

\textbf{{Margins.jl}} provides the \texttt{second\_differences()}
function family to compute these quantities with proper delta-method
standard errors. Two computational approaches are available: (1)
discrete contrasts comparing pre-computed AMEs across moderator levels,
and (2) local derivatives computing \(\partial\text{AME}/\partial Z\)
via finite differences, evaluated at the sample mean of \(Z\) by
default.

Continuous \(\times\) Categorical: Does the effect of a continuous
variable differ across groups?

\begin{Shaded}
\begin{Highlighting}[]
\CommentTok{\# Effect of income on purchase probability by education level}
\NormalTok{model }\OperatorTok{=} \FunctionTok{glm}\NormalTok{(}\PreprocessorTok{@formula}\NormalTok{(purchase }\OperatorTok{\textasciitilde{}}\NormalTok{ income }\OperatorTok{*}\NormalTok{ education }\OperatorTok{+}\NormalTok{ age), data, }\FunctionTok{Binomial}\NormalTok{())}

\CommentTok{\# Compute AMEs across education levels}
\NormalTok{ames }\OperatorTok{=} \FunctionTok{population\_margins}\NormalTok{(model, data;}
\NormalTok{    scenarios}\OperatorTok{=}\NormalTok{(education}\OperatorTok{=}\NormalTok{[}\StringTok{"hs"}\NormalTok{, }\StringTok{"college"}\NormalTok{, }\StringTok{"grad"}\NormalTok{],),}
    \KeywordTok{type}\OperatorTok{=:}\NormalTok{effects)}

\CommentTok{\# Test whether income effect differs by education}
\NormalTok{sd }\OperatorTok{=} \FunctionTok{second\_differences}\NormalTok{(ames, }\OperatorTok{:}\NormalTok{income, }\OperatorTok{:}\NormalTok{education, }\FunctionTok{vcov}\NormalTok{(model))}
\end{Highlighting}
\end{Shaded}

Continuous \(\times\) Continuous: How does one continuous effect change
with another continuous variable?

\begin{Shaded}
\begin{Highlighting}[]
\CommentTok{\# Effect of price on demand varies with household income}
\NormalTok{model }\OperatorTok{=} \FunctionTok{lm}\NormalTok{(}\PreprocessorTok{@formula}\NormalTok{(quantity }\OperatorTok{\textasciitilde{}}\NormalTok{ price }\OperatorTok{*}\NormalTok{ income }\OperatorTok{+}\NormalTok{ controls), data)}

\CommentTok{\# Local derivative: how does price effect change with income?}
\NormalTok{sd }\OperatorTok{=} \FunctionTok{second\_differences\_at}\NormalTok{(model, data, }\OperatorTok{:}\NormalTok{price, }\OperatorTok{:}\NormalTok{income, }\FunctionTok{vcov}\NormalTok{(model);}
\NormalTok{    at}\OperatorTok{=}\NormalTok{[}\FloatTok{30000}\NormalTok{, }\FloatTok{50000}\NormalTok{, }\FloatTok{100000}\NormalTok{])  }\CommentTok{\# Evaluate at specific income levels}
\end{Highlighting}
\end{Shaded}

This computation uses symmetric finite differences rather than automatic
differentiation; see Appendix A for details on step size selection and
numerical stability.

Categorical \(\times\) Categorical: Does the effect of one categorical
variable depend on another?

\begin{Shaded}
\begin{Highlighting}[]
\CommentTok{\# Employment effect on wellbeing varies by region}
\NormalTok{model }\OperatorTok{=} \FunctionTok{lm}\NormalTok{(}\PreprocessorTok{@formula}\NormalTok{(wellbeing }\OperatorTok{\textasciitilde{}}\NormalTok{ employed }\OperatorTok{*}\NormalTok{ region }\OperatorTok{+}\NormalTok{ age), data)}

\CommentTok{\# Compute AMEs across regions}
\NormalTok{ames }\OperatorTok{=} \FunctionTok{population\_margins}\NormalTok{(model, data;}
\NormalTok{    scenarios}\OperatorTok{=}\NormalTok{(region}\OperatorTok{=}\NormalTok{[}\StringTok{"north"}\NormalTok{, }\StringTok{"south"}\NormalTok{, }\StringTok{"east"}\NormalTok{, }\StringTok{"west"}\NormalTok{],),}
    \KeywordTok{type}\OperatorTok{=:}\NormalTok{effects)}

\CommentTok{\# All pairwise comparisons of employment effect across regions}
\NormalTok{sd }\OperatorTok{=} \FunctionTok{second\_differences\_all\_contrasts}\NormalTok{(ames, }\OperatorTok{:}\NormalTok{employed, }\OperatorTok{:}\NormalTok{region, }\FunctionTok{vcov}\NormalTok{(model))}
\end{Highlighting}
\end{Shaded}

Standard errors for second differences are computed via the delta
method, properly accounting for covariance between the underlying
marginal effect estimates:

\[\text{SE}(\Delta) = \sqrt{(g_2 - g_1)^T \Sigma (g_2 - g_1)}\]

where \(g_1\) and \(g_2\) are the gradient vectors of each marginal
effect with respect to model parameters, and \(\Sigma\) is the parameter
covariance matrix.

\subsection{Integration with Julia
Ecosystem}\label{integration-with-julia-ecosystem}

The integration between \textbf{{Margins.jl}} and {Julia}'s statistical
ecosystem begins at the formula layer. {StatsModels.jl} (Kleinschmidt et
al. 2024) provides the formula interface that both packages build upon.
\textbf{{FormulaCompiler.jl}} decomposes \texttt{FormulaTerm} objects
into atomic operations, handling all standard term types including
continuous variables, categorical terms with their contrast matrices,
function applications (e.g., \texttt{log()}, \texttt{sqrt()}), and
multi-way interactions. The {Tables.jl} (Quinn 2024) interface enables
the zero-allocation evaluation architecture: data is converted to
columnar \texttt{NamedTuple} format via \texttt{Tables.columntable()},
allowing type-stable column access where each load operation resolves to
a direct property access at compile time.

For model fitting, \textbf{{Margins.jl}} extracts the necessary
components through standard {StatsAPI} methods. From {GLM.jl} (Bates et
al. 2024) models, coefficients are obtained via \texttt{coef()},
parameter covariance via \texttt{vcov()}, and link functions via
\texttt{Link(model)}. All standard GLM link functions are
supported---identity, log, logit, probit, complementary log-log,
cauchit, inverse, and square root---with both first and second
derivatives implemented for delta-method variance computation. For
{MixedModels.jl} (Bates, Alday, and Kokandakar 2025),
\textbf{{FormulaCompiler.jl}} extracts the fixed effects formula by
stripping random effects terms, enabling marginal effects computation
for both \texttt{LinearMixedModel} and
\texttt{GeneralizedLinearMixedModel} objects. As noted in the Future
Work section, marginal effects are currently computed for fixed effects
only.

Categorical variables receive special treatment through integration with
{CategoricalArrays.jl} (Bouchet-Valat, Kamiński, and Quinn 2024).
Variables are automatically detected via their \texttt{CategoricalArray}
type, and contrast matrices are accessed directly from
{StatsModels.jl}'s contrast coding specifications---supporting all
standard schemes including dummy coding, effects coding, and Helmert
coding. For profile-based analysis, \textbf{{FormulaCompiler.jl}}
provides \texttt{CategoricalMixture} types that represent weighted
combinations of categorical levels, enabling representative estimates
that avoid arbitrary reference level selection. Boolean variables are
handled analogously, with the probability of \texttt{true} computed from
sample frequencies for mixture-based evaluation.

The data pipeline leverages {Julia}'s interface abstractions for
flexibility. Input data can be any {Tables.jl}-compatible
format---{DataFrames.jl} (Bouchet-Valat and Kamiński 2023),
\texttt{NamedTuple}, or custom table types---with automatic conversion
to the columnar format required for efficient evaluation. Output follows
the same pattern: result types implement the {Tables.jl} interface
(\texttt{Tables.istable}, \texttt{Tables.rows}, \texttt{Tables.schema}),
enabling direct conversion to DataFrames with configurable output
formats. This bidirectional {Tables.jl} integration allows
\textbf{{Margins.jl}} results to flow naturally into downstream analysis
pipelines.

Several additional packages extend the core functionality.
{StandardizedPredictors.jl} (Kleinschmidt and Alday 2024) enables
automatic back-transformation: when models are fit with z-scored
predictors, \textbf{{FormulaCompiler.jl}} recognizes
\texttt{ZScoredTerm} objects and reports marginal effects on the
original variable scale. Support for other standardization types
(\emph{e.g.}, \texttt{CenteredTerm}) is planned for future releases.
{CovarianceMatrices.jl} (Ragusa 2024) provides heteroskedasticity-robust
(HC0--HC3) and cluster-robust (CR0/CR1) standard errors through the
\texttt{vcov} parameter (see the Robust Standard Errors section). The
automatic differentiation underlying derivative computation uses
{ForwardDiff.jl} (Revels, Lubin, and Papamarkou 2016), with dual numbers
pre-converted at construction time to eliminate per-access type
conversion overhead---a key factor in achieving zero-allocation
evaluation.

For models fit with standardized predictors, marginal effects are
automatically reported on the original scale:

\begin{Shaded}
\begin{Highlighting}[]
\NormalTok{model }\OperatorTok{=} \FunctionTok{lm}\NormalTok{(}\PreprocessorTok{@formula}\NormalTok{(y }\OperatorTok{\textasciitilde{}}\NormalTok{ x }\OperatorTok{+}\NormalTok{ z), data, contrasts}\OperatorTok{=}\FunctionTok{Dict}\NormalTok{(}\OperatorTok{:}\NormalTok{x }\OperatorTok{=\textgreater{}} \FunctionTok{ZScore}\NormalTok{()))}
\FunctionTok{population\_margins}\NormalTok{(model, data)  }\CommentTok{\# Effects per unit of raw x, not per SD}
\end{Highlighting}
\end{Shaded}

\subsection{Robust Standard Errors}\label{robust-standard-errors}

\textbf{{Margins.jl}} supports heteroskedasticity-robust and
cluster-robust standard errors through integration with
{CovarianceMatrices.jl} (Zeileis 2004, 2006). The \texttt{vcov}
parameter accepts any covariance estimator, enabling robust inference
without modifying the marginal effects computation.

Heteroskedasticity-robust standard errors:

\begin{Shaded}
\begin{Highlighting}[]
\ImportTok{using} \BuiltInTok{CovarianceMatrices}

\CommentTok{\# HC1 (most common)}
\FunctionTok{population\_margins}\NormalTok{(model, data; vcov}\OperatorTok{=}\FunctionTok{HC1}\NormalTok{())}

\CommentTok{\# HC3 (more conservative)}
\FunctionTok{population\_margins}\NormalTok{(model, data; vcov}\OperatorTok{=}\FunctionTok{HC3}\NormalTok{())}
\end{Highlighting}
\end{Shaded}

Cluster-robust standard errors:

\begin{Shaded}
\begin{Highlighting}[]
\CommentTok{\# Single{-}level clustering by firm}
\FunctionTok{population\_margins}\NormalTok{(model, data; vcov}\OperatorTok{=}\FunctionTok{CR0}\NormalTok{(data.firm\_id))}
\end{Highlighting}
\end{Shaded}

The implementation uses the delta method to propagate robust covariance
estimates through the marginal effects transformation. For a marginal
effect \(\theta = g(\beta)\) with gradient \(\nabla_\beta g\), the
robust standard error is:

\[\text{SE}_{\text{robust}}(\hat{\theta}) = \sqrt{\nabla_\beta g(\hat{\beta})^T \, \hat{\Sigma}_{\text{robust}} \, \nabla_\beta g(\hat{\beta})}\]

where \(\hat{\Sigma}_{\text{robust}}\) is the sandwich estimator
(HC0-HC3) or cluster-robust estimator (CR0/CR1). This approach ensures
that robust inference for marginal effects inherits the same theoretical
properties as robust inference for the underlying model coefficients.

\section{FormulaCompiler.jl: Computational
Foundation}\label{formulacompiler.jl-computational-foundation}

The performance characteristics of \textbf{{Margins.jl}} derive from
\textbf{{FormulaCompiler.jl}}, which transforms statistical formulas
into zero-allocation evaluators through a position-mapped compilation
strategy. ``Zero-allocation'' refers to per-row evaluation: construction
incurs one-time \(O(p)\) allocations for buffers, but subsequent
evaluation adds no allocations that scale with dataset size \(n\)---the
key to avoiding the memory constraints that limit existing
implementations. This section describes the key architectural elements
that enable efficient marginal effects computation.

\subsection{Position-Mapped
Compilation}\label{position-mapped-compilation}

Traditional formula systems construct the full \(O(n \times p)\) design
matrix before any computation. For marginal effects, this creates a
bottleneck: computing average marginal effects requires evaluating the
formula at every observation, and delta-method standard errors require
accumulating gradients across all \(n\) rows. Materializing the full
matrix, or the \(n \times p\) Jacobian may become prohibitive at scale.

\textbf{{FormulaCompiler.jl}} compiles {StatsModels.jl}-compatible
formulas into fully-typed evaluator structs where every operation's
output position is determined at compile time. The compilation process
begins by decomposing each formula element---variable references,
transformations, and interactions---into atomic operations. Each term
\(j\) is then assigned a fixed output position
\(\text{Position}_j \in \{1,\ldots,p\}\), creating a static mapping from
formula structure to memory layout.

\textbf{{FormulaCompiler.jl}} builds directly on {StatsModels.jl}'s
Wilkinson notation formula system. The package extracts the
\texttt{FormulaTerm} from the fitted model's schema (its stored formula
and contrast specifications) and traverses the term tree,
\texttt{InteractionTerm}, \texttt{FunctionTerm},
\texttt{CategoricalTerm} and other term types, to generate the operation
sequence. Contrast matrices for categorical variables are obtained from
{StatsModels.jl}'s contrast specifications, ensuring consistency with
the original model fit. This reuse strategy means that any formula
accepted by {GLM.jl} or {MixedModels.jl} works identically with
\textbf{{FormulaCompiler.jl}}.

The compiled formula is represented as a fully-typed struct where all
structural information is encoded in type parameters:

\begin{Shaded}
\begin{Highlighting}[]
\KeywordTok{struct}\NormalTok{ UnifiedCompiled\{T, OpsTuple, ScratchSize, OutputSize\}}
\NormalTok{    ops}\OperatorTok{::}\DataTypeTok{OpsTuple}
\NormalTok{    scratch}\OperatorTok{::}\DataTypeTok{Vector\{T\}}
\KeywordTok{end}
\end{Highlighting}
\end{Shaded}

The type parameters encode the complete evaluation strategy: \texttt{T}
is the element type (typically \texttt{Float64}, or \texttt{Dual} for
automatic differentiation), \texttt{OpsTuple} contains the sequence of
typed operations, and \texttt{ScratchSize} and \texttt{OutputSize} are
compile-time constants specifying buffer dimensions.

Operations are encoded as subtypes of \texttt{AbstractOp}, with
positions embedded in type parameters:

\begin{Shaded}
\begin{Highlighting}[]
\KeywordTok{struct}\NormalTok{ LoadOp\{Column, OutPos\} }\OperatorTok{\textless{}:}\DataTypeTok{ AbstractOp }\KeywordTok{end}
\KeywordTok{struct}\NormalTok{ BinaryOp\{Func, InPos1, InPos2, OutPos\} }\OperatorTok{\textless{}:}\DataTypeTok{ AbstractOp }\KeywordTok{end}
\KeywordTok{struct}\NormalTok{ ContrastOp\{Column, OutPositions\}}
\NormalTok{    contrast\_matrix}\OperatorTok{::}\DataTypeTok{Matrix\{Float64\}}
\KeywordTok{end}
\end{Highlighting}
\end{Shaded}

For example, \texttt{LoadOp\{:x,\ 3\}} loads column \texttt{:x} into
scratch position 3, and \texttt{BinaryOp\{:*,\ 1,\ 2,\ 4\}} multiplies
positions 1 and 2, storing the result in position 4. This encoding
embeds all indexing decisions in type parameters, enabling the Julia
compiler to specialize evaluation code for each unique formula
structure. At runtime, the evaluator simply executes a fixed sequence of
operations with no branching or dynamic lookups.

{Julia}'s tuple specialization heuristics can fail for large operation
sequences, causing performance degradation.
\textbf{{FormulaCompiler.jl}} uses a hybrid dispatch strategy: formulas
with \(\leq 10\) operations use recursion, while larger formulas use
\texttt{@generated} functions.\footnote{\texttt{@generated} functions
  are a Julia feature where the function body itself is computed at
  compile time based on the types of the arguments. This enables the
  compiler to produce specialized, unrolled code for each unique formula
  structure.} \texttt{@generated} refers to {Julia}'s metaprogramming
capacity for compile-time code generation (Bezanson et al. 2017), which
produces specialized code by inspecting the operation types at compile
time, unrolling the entire operation sequence into a single function
body with no loops or recursion. This ensures zero allocations
regardless of formula complexity---a formula with 100+ terms achieves
identical per-row performance to simple formulas.

\begin{figure}

\centering{

\pandocbounded{\includegraphics[keepaspectratio]{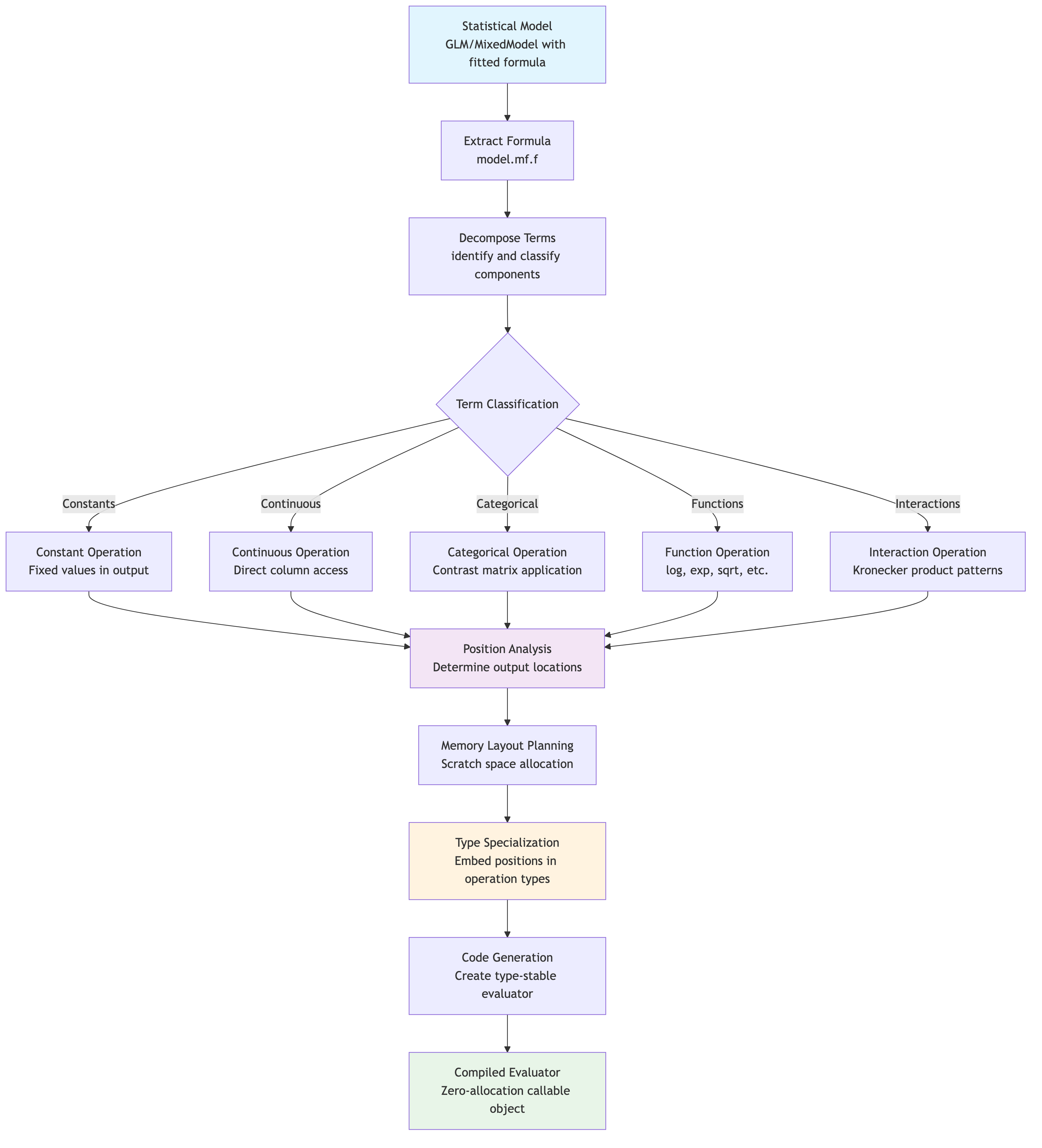}}

}

\caption{\label{fig-compilation}Compilation pipeline:
{FormulaCompiler.jl} transforms {StatsModels.jl} formulas through four
stages---(1) term decomposition into atomic operations, (2) position
mapping to fixed output indices, (3) type encoding of operations in
{Julia}'s type parameters, and (4) code generation with buffer
preallocation---to produce zero-allocation evaluators.}

\end{figure}

A fixed-size scratch vector is preallocated and reused across
evaluations, with \texttt{@inline} annotations and bounds-check
elimination ensuring zero overhead per operation. The compile-once,
evaluate-many pattern is ideal for marginal effects computation, where
the same formula must be evaluated at every observation.
Figure~\ref{fig-compilation} illustrates this pipeline from formula
input to compiled evaluator.

\begin{Shaded}
\begin{Highlighting}[]
\CommentTok{\# Example: formula to compiled evaluator}
\NormalTok{model }\OperatorTok{=} \FunctionTok{lm}\NormalTok{(}\PreprocessorTok{@formula}\NormalTok{(y }\OperatorTok{\textasciitilde{}}\NormalTok{ x1 }\OperatorTok{+}\NormalTok{ x2 }\OperatorTok{*}\NormalTok{ group), data)}
\NormalTok{compiled }\OperatorTok{=} \FunctionTok{compile\_formula}\NormalTok{(model, data)  }\CommentTok{\# Stage 4 output}
\NormalTok{output }\OperatorTok{=} \FunctionTok{Vector}\DataTypeTok{\{Float64\}}\NormalTok{(}\ConstantTok{undef}\NormalTok{, }\FunctionTok{width}\NormalTok{(compiled))}
\FunctionTok{compiled}\NormalTok{(output, data, }\FloatTok{1}\NormalTok{)  }\CommentTok{\# Zero{-}allocation row evaluation}
\end{Highlighting}
\end{Shaded}

\subsection{Zero-Allocation
Evaluation}\label{zero-allocation-evaluation}

Given the compiled evaluator and a columnar data interface ({Tables.jl};
Quinn (2024)), row evaluation proceeds in place. The formula is compiled
once, and the resulting evaluator can be called repeatedly with zero
heap allocations:

\begin{Shaded}
\begin{Highlighting}[]
\NormalTok{compiled }\OperatorTok{=} \FunctionTok{compile\_formula}\NormalTok{(model, data)}
\NormalTok{row }\OperatorTok{=} \FunctionTok{Vector}\DataTypeTok{\{Float64\}}\NormalTok{(}\ConstantTok{undef}\NormalTok{, }\FunctionTok{length}\NormalTok{(compiled))}

\ControlFlowTok{for}\NormalTok{ i }\KeywordTok{in} \FloatTok{1}\OperatorTok{:}\FunctionTok{nrow}\NormalTok{(data)}
    \FunctionTok{compiled}\NormalTok{(row, data, i)}
\ControlFlowTok{end}
\end{Highlighting}
\end{Shaded}

Each iteration allocates zero bytes after warmup. Core evaluation
requires only \(O(p)\) memory through preallocated buffers, with zero
heap allocations after warmup; scenario analysis adds only \(O(1)\)
overhead through counterfactual vector types that present modified
values to the evaluator while the underlying data remains unchanged.

\begin{figure}

\centering{

\pandocbounded{\includegraphics[keepaspectratio]{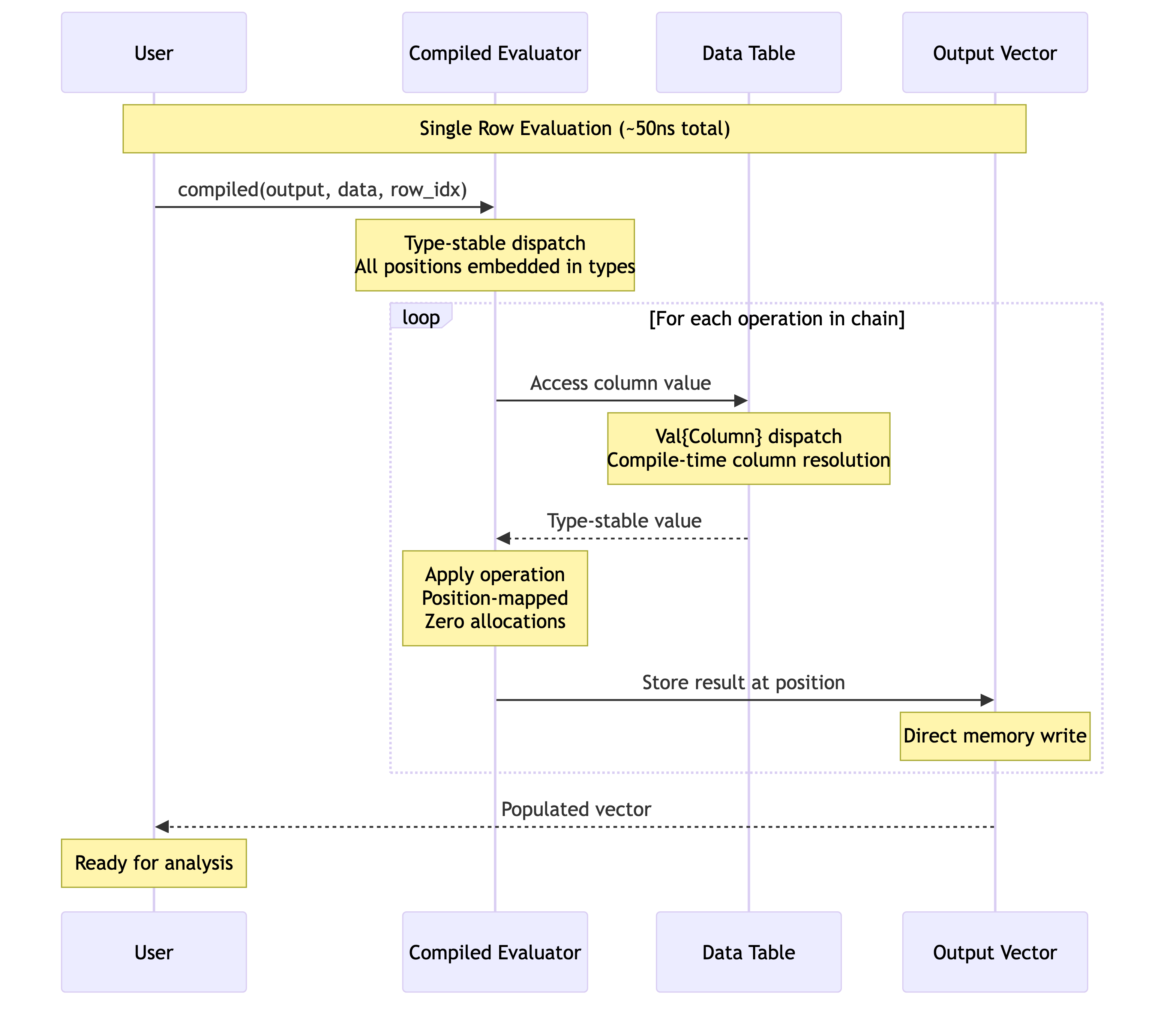}}

}

\caption{\label{fig-execution}Sequence diagram showing zero-allocation
row evaluation: type-stable dispatch resolves column access at compile
time, enabling direct memory writes without intermediate allocations.}

\end{figure}

This is fundamentally different from traditional implementations that
materialize \(O(n \times p)\) design matrices:
\textbf{{FormulaCompiler.jl}} never constructs the full matrix, instead
evaluating each row on demand. Figure~\ref{fig-execution} shows the
sequence of operations during row evaluation, highlighting type-stable
dispatch and direct memory writes.

\begin{Shaded}
\begin{Highlighting}[]
\CommentTok{\# Zero{-}allocation evaluation loop}
\NormalTok{compiled }\OperatorTok{=} \FunctionTok{compile\_formula}\NormalTok{(model, data)}
\NormalTok{row }\OperatorTok{=} \FunctionTok{Vector}\DataTypeTok{\{Float64\}}\NormalTok{(}\ConstantTok{undef}\NormalTok{, }\FunctionTok{width}\NormalTok{(compiled))}

\ControlFlowTok{for}\NormalTok{ i }\KeywordTok{in} \FloatTok{1}\OperatorTok{:}\FunctionTok{nrow}\NormalTok{(data)}
    \FunctionTok{compiled}\NormalTok{(row, data, i)  }\CommentTok{\# Updates row in{-}place, zero allocations}
\ControlFlowTok{end}
\end{Highlighting}
\end{Shaded}

\subsection{Automatic Differentiation}\label{automatic-differentiation}

Marginal effects require derivatives of the model row with respect to
input variables. \textbf{{FormulaCompiler.jl}} provides this through a
manually orchestrated automatic differentiation (AD) system using
{ForwardDiff.jl} (Revels, Lubin, and Papamarkou 2016) dual numbers,
which carry both a value and its derivative, enabling exact derivative
computation through the chain rule.

A practical challenge in applying automatic differentiation to
statistical formulas is type conversion overhead. Standard AD approaches
convert \texttt{Float64} data to dual numbers on every access, creating
allocations that scale with formula complexity.
\textbf{{FormulaCompiler.jl}} implements a pre-conversion strategy that
eliminates this overhead through two phases:

\begin{enumerate}
\def\labelenumi{\arabic{enumi}.}
\item
  \textbf{Construction-time pre-conversion}: During engine construction,
  data columns are converted once from \texttt{Float64} to
  \texttt{Dual\{T,Float64,k\}} for \(k\) differentiation variables. This
  amortized cost occurs before any per-row computation.
\item
  \textbf{Type-homogeneous evaluation}: During derivative computation,
  the evaluation chain maintains dual types throughout---no runtime
  conversions occur. Variables are seeded with identity partials, the
  compiled formula executes on dual-typed data, and partial derivatives
  are extracted directly from results.
\end{enumerate}

Rather than using {ForwardDiff.jl}'s general-purpose \texttt{jacobian!}
function (which incurs allocation overhead for flexibility),
\textbf{{FormulaCompiler.jl}} implements direct seeding, in-place buffer
updates, single-pass evaluation, and loop-based extraction. This
achieves ForwardDiff's mathematical correctness with custom
zero-allocation orchestration.

The result: traditional AD incurs
\(O(\text{accesses} \times \text{conversions})\) runtime allocations,
while this approach achieves zero runtime allocations independent of
formula complexity. A derivative evaluator is constructed for the
variables of interest, and marginal effects are computed in place:

\begin{Shaded}
\begin{Highlighting}[]
\NormalTok{de }\OperatorTok{=} \FunctionTok{build\_derivative\_evaluator}\NormalTok{(compiled, data; vars}\OperatorTok{=}\NormalTok{[}\OperatorTok{:}\NormalTok{x1, }\OperatorTok{:}\NormalTok{x2])}

\NormalTok{g }\OperatorTok{=} \FunctionTok{Vector}\DataTypeTok{\{Float64\}}\NormalTok{(}\ConstantTok{undef}\NormalTok{, }\FloatTok{2}\NormalTok{)}
\FunctionTok{marginal\_effects\_mu!}\NormalTok{(g, de, }\FunctionTok{coef}\NormalTok{(model), row\_idx; link}\OperatorTok{=}\FunctionTok{LogitLink}\NormalTok{())}
\end{Highlighting}
\end{Shaded}

Derivative computation uses specialized evaluators that maintain type
stability throughout. \texttt{ADEvaluator} pre-converts data columns to
dual numbers at construction time, enabling zero-allocation Jacobian
computation. \texttt{FDEvaluator} provides finite difference derivatives
as an alternative to automatic differentiation. Both evaluator types
hold pre-allocated buffers sized to the formula's output width, ensuring
that derivative computation adds no per-row allocations beyond the base
formula evaluation.

\textbf{{FormulaCompiler.jl}} supports marginal effects computation for
all standard link functions. The derivative of the inverse link
function, \(d\mu/d\eta\), is required to transform linear predictor
effects to the response scale via the chain rule. Supported links
include identity, log, logit, probit, complementary log-log (cloglog),
cauchit, inverse, sqrt, and inverse-square. For binary outcomes, the
logit and probit links are most common; the log link is standard for
count data (Poisson, negative binomial); identity is used for linear
models. Second derivatives are computed automatically for delta-method
standard errors when effects are evaluated on the response scale.

\subsection{Enabling Margins.jl}\label{enabling-margins.jl}

\textbf{{FormulaCompiler.jl}} exposes a small set of primitives that
\textbf{{Margins.jl}} composes to implement the full statistical
interface: \texttt{compile\_formula()} for one-time formula compilation,
\texttt{continuous\_variables()} for automatic detection of variable
types, \texttt{marginal\_effects\_eta!()} and
\texttt{marginal\_effects\_mu!()} for per-row effect computation on the
linear predictor and response scales respectively,
\texttt{accumulate\_ame\_gradient!()} for efficient gradient
accumulation across observations, and \texttt{delta\_method\_se()} for
standard error computation from gradients and the parameter covariance
matrix.

\textbf{{Margins.jl}} uses marker types to specialize computation at
compile time:

\begin{Shaded}
\begin{Highlighting}[]
\KeywordTok{abstract type}\NormalTok{ MarginsUsage }\KeywordTok{end}
\KeywordTok{struct}\NormalTok{ PopulationUsage }\OperatorTok{\textless{}:}\DataTypeTok{ MarginsUsage }\KeywordTok{end}
\KeywordTok{struct}\NormalTok{ ProfileUsage }\OperatorTok{\textless{}:}\DataTypeTok{ MarginsUsage }\KeywordTok{end}

\KeywordTok{abstract type}\NormalTok{ DerivativeSupport }\KeywordTok{end}
\KeywordTok{struct}\NormalTok{ HasDerivatives }\OperatorTok{\textless{}:}\DataTypeTok{ DerivativeSupport}
\NormalTok{    evaluator}\OperatorTok{::}\DataTypeTok{AbstractDerivativeEvaluator}
\KeywordTok{end}
\KeywordTok{struct}\NormalTok{ NoDerivatives }\OperatorTok{\textless{}:}\DataTypeTok{ DerivativeSupport }\KeywordTok{end}
\end{Highlighting}
\end{Shaded}

These marker types correspond to the two API entry points:
\texttt{PopulationUsage} for \texttt{population\_margins()} and
\texttt{ProfileUsage} for \texttt{profile\_margins()}. They flow into
the main computation engine:

\begin{Shaded}
\begin{Highlighting}[]
\KeywordTok{struct}\NormalTok{ MarginsEngine\{L}\OperatorTok{\textless{}:}\DataTypeTok{GLM.Link}\NormalTok{, U}\OperatorTok{\textless{}:}\DataTypeTok{MarginsUsage}\NormalTok{, D}\OperatorTok{\textless{}:}\DataTypeTok{DerivativeSupport}\NormalTok{, C\}}
\NormalTok{    compiled}\OperatorTok{::}\DataTypeTok{C}
\NormalTok{    link}\OperatorTok{::}\DataTypeTok{L}
    \CommentTok{\# ... pre{-}allocated buffers}
\KeywordTok{end}
\end{Highlighting}
\end{Shaded}

The type parameters enable {Julia}'s multiple dispatch to select
optimized code paths: \texttt{PopulationUsage} triggers row-streaming
algorithms with minimal memory footprint, while \texttt{ProfileUsage}
optimizes for evaluating multiple covariate profiles. The
\texttt{compiled} field holds a \texttt{UnifiedCompiled} evaluator from
\textbf{{FormulaCompiler.jl}}, forming the bridge between the
statistical interface and the computational foundation. This design
enables \textbf{{Margins.jl}} to compute population-averaged quantities
by iterating efficiently over observations without the memory overhead
that limits existing implementations.

\subsection{Scenario System}\label{scenario-system}

For counterfactual analysis, \textbf{{FormulaCompiler.jl}} provides a
scenario system that overrides variable values without copying data,
achieving \(O(1)\) memory overhead per scenario. This functionality
parallels {Stata}'s \texttt{margins} command with \texttt{at()} options
(Williams 2012). For example, to compute effects under the
counterfactual where \texttt{x1\ =\ 1.0} for all observations:

\begin{Shaded}
\begin{Highlighting}[]
\NormalTok{scenario\_data }\OperatorTok{=} \FunctionTok{create\_scenario}\NormalTok{(data; x1}\OperatorTok{=}\FloatTok{1.0}\NormalTok{)}

\FunctionTok{compiled}\NormalTok{(row, scenario\_data, i)}
\end{Highlighting}
\end{Shaded}

Evaluation proceeds identically to the original data, with \(O(1)\)
overhead. Counterfactual vector types present modified values to the
evaluator while the underlying data remains unchanged, enabling
counterfactual evaluation with minimal memory overhead. Single-row
perturbations for derivative computation use type-specialized vectors:

\begin{Shaded}
\begin{Highlighting}[]
\CommentTok{\# Numeric override}
\FunctionTok{NumericCounterfactualVector}\NormalTok{(data.x, }\FloatTok{1.0}\NormalTok{)  }\CommentTok{\# x = 1.0 for all}

\CommentTok{\# Categorical override}
\FunctionTok{CategoricalCounterfactualVector}\NormalTok{(data.group, }\StringTok{"B"}\NormalTok{)  }\CommentTok{\# group = "B" for all}

\CommentTok{\# Boolean override}
\FunctionTok{BoolCounterfactualVector}\NormalTok{(data.treated, }\ConstantTok{true}\NormalTok{)  }\CommentTok{\# treated = true for all}
\end{Highlighting}
\end{Shaded}

These vectors present modified values while delegating to base data
elsewhere, enabling derivative computation at counterfactual values
without copying the underlying dataset.

\section{Performance Benchmarks}\label{performance-benchmarks}

This section presents performance comparisons demonstrating the
practical impact of the architectural approach. We first compare
\textbf{{Margins.jl}} against {R}'s state-of-the-art {marginaleffects}
package, then present micro-benchmarks validating
\textbf{{FormulaCompiler.jl}}'s zero-allocation guarantees.
Table~\ref{tbl-micro-benchmarks} and Table~\ref{tbl-case-studies}
provide complete specifications for benchmark formulas and dataset
sizes.

\subsection{Margins.jl vs R
marginaleffects}\label{margins.jl-vs-r-marginaleffects}

We benchmark \textbf{{Margins.jl}} against {R}'s {marginaleffects}
package (Arel-Bundock, Greifer, and Heiss 2024) on a realistic logistic
regression model with 65 parameters, including interactions between
continuous and categorical variables. Both packages compute identical
quantities using equivalent statistical methodology.
Table~\ref{tbl-bench-standard} presents results for the standard
dataset, and Table~\ref{tbl-bench-large} shows performance at scale.

The following code demonstrates how each package computes average
marginal effects:

\textbf{Julia}

\begin{Shaded}
\begin{Highlighting}[]
\ImportTok{using} \BuiltInTok{Margins}\NormalTok{, }\BuiltInTok{GLM}\NormalTok{, }\BuiltInTok{DataFrames}

\NormalTok{model }\OperatorTok{=} \FunctionTok{glm}\NormalTok{(}\PreprocessorTok{@formula}\NormalTok{(y }\OperatorTok{\textasciitilde{}}\NormalTok{ x1 }\OperatorTok{*}\NormalTok{ x2 }\OperatorTok{+}\NormalTok{ x3 }\OperatorTok{*}\NormalTok{ cat1), data, }\FunctionTok{Binomial}\NormalTok{())}
\NormalTok{result }\OperatorTok{=} \FunctionTok{population\_margins}\NormalTok{(model; }\KeywordTok{type}\OperatorTok{=:}\NormalTok{effects)  }\CommentTok{\# AME for all variables}
\end{Highlighting}
\end{Shaded}

\textbf{R}

\begin{Shaded}
\begin{Highlighting}[]
\FunctionTok{library}\NormalTok{(marginaleffects)}

\NormalTok{model }\OtherTok{\textless{}{-}} \FunctionTok{glm}\NormalTok{(y }\SpecialCharTok{\textasciitilde{}}\NormalTok{ x1 }\SpecialCharTok{*}\NormalTok{ x2 }\SpecialCharTok{+}\NormalTok{ x3 }\SpecialCharTok{*}\NormalTok{ cat1, }\AttributeTok{data=}\NormalTok{data, }\AttributeTok{family=}\FunctionTok{binomial}\NormalTok{())}
\NormalTok{result }\OtherTok{\textless{}{-}} \FunctionTok{avg\_slopes}\NormalTok{(model)  }\CommentTok{\# AME for all variables}
\end{Highlighting}
\end{Shaded}

Standard dataset (5,000 observations):

\begin{longtable}[]{@{}
  >{\raggedright\arraybackslash}p{(\linewidth - 12\tabcolsep) * \real{0.1196}}
  >{\raggedright\arraybackslash}p{(\linewidth - 12\tabcolsep) * \real{0.1848}}
  >{\raggedright\arraybackslash}p{(\linewidth - 12\tabcolsep) * \real{0.2283}}
  >{\raggedright\arraybackslash}p{(\linewidth - 12\tabcolsep) * \real{0.0978}}
  >{\raggedright\arraybackslash}p{(\linewidth - 12\tabcolsep) * \real{0.1304}}
  >{\raggedright\arraybackslash}p{(\linewidth - 12\tabcolsep) * \real{0.0870}}
  >{\raggedright\arraybackslash}p{(\linewidth - 12\tabcolsep) * \real{0.1522}}@{}}
\caption{Performance comparison on standard dataset (5,000 observations,
65-parameter logistic
regression).}\label{tbl-bench-standard}\tabularnewline
\toprule\noalign{}
\begin{minipage}[b]{\linewidth}\raggedright
Operation
\end{minipage} & \begin{minipage}[b]{\linewidth}\raggedright
Margins.jl (ms)
\end{minipage} & \begin{minipage}[b]{\linewidth}\raggedright
marginaleffects (ms)
\end{minipage} & \begin{minipage}[b]{\linewidth}\raggedright
Speedup
\end{minipage} & \begin{minipage}[b]{\linewidth}\raggedright
Julia (MB)
\end{minipage} & \begin{minipage}[b]{\linewidth}\raggedright
R (MB)
\end{minipage} & \begin{minipage}[b]{\linewidth}\raggedright
Memory Ratio
\end{minipage} \\
\midrule\noalign{}
\endfirsthead
\toprule\noalign{}
\begin{minipage}[b]{\linewidth}\raggedright
Operation
\end{minipage} & \begin{minipage}[b]{\linewidth}\raggedright
Margins.jl (ms)
\end{minipage} & \begin{minipage}[b]{\linewidth}\raggedright
marginaleffects (ms)
\end{minipage} & \begin{minipage}[b]{\linewidth}\raggedright
Speedup
\end{minipage} & \begin{minipage}[b]{\linewidth}\raggedright
Julia (MB)
\end{minipage} & \begin{minipage}[b]{\linewidth}\raggedright
R (MB)
\end{minipage} & \begin{minipage}[b]{\linewidth}\raggedright
Memory Ratio
\end{minipage} \\
\midrule\noalign{}
\endhead
\bottomrule\noalign{}
\endlastfoot
APM (Adjusted Predictions at Profiles) & 0.70 & 76 & 109x & 0.93 & 12.8
& 14x \\
MEM (Marginal Effects at Profiles) & 2.1 & 6413 & 3124x & 1.9 & 719 &
383x \\
AAP (Average Adjusted Predictions) & 0.59 & 154 & 261x & 0.46 & 495 &
1072x \\
AME (all variables) & 57 & 2974 & 52x & 74.9 & 14640 & 196x \\
AME (single variable) & 1.7 & 299 & 173x & 0.46 & 505 & 1092x \\
AME (scenario) & 21 & 294 & 14x & 10.8 & 34.3 & 3x \\
\textbf{Average} & --- & --- & \textbf{622x} & --- & --- &
\textbf{460x} \\
\end{longtable}

Large dataset (500,000 observations):

\begin{longtable}[]{@{}
  >{\raggedright\arraybackslash}p{(\linewidth - 4\tabcolsep) * \real{0.2821}}
  >{\raggedright\arraybackslash}p{(\linewidth - 4\tabcolsep) * \real{0.2821}}
  >{\raggedright\arraybackslash}p{(\linewidth - 4\tabcolsep) * \real{0.4359}}@{}}
\caption{Performance comparison on large dataset (500,000 observations).
R's marginaleffects fails on AME due to memory
exhaustion.}\label{tbl-bench-large}\tabularnewline
\toprule\noalign{}
\begin{minipage}[b]{\linewidth}\raggedright
Operation
\end{minipage} & \begin{minipage}[b]{\linewidth}\raggedright
Margins.jl
\end{minipage} & \begin{minipage}[b]{\linewidth}\raggedright
marginaleffects
\end{minipage} \\
\midrule\noalign{}
\endfirsthead
\toprule\noalign{}
\begin{minipage}[b]{\linewidth}\raggedright
Operation
\end{minipage} & \begin{minipage}[b]{\linewidth}\raggedright
Margins.jl
\end{minipage} & \begin{minipage}[b]{\linewidth}\raggedright
marginaleffects
\end{minipage} \\
\midrule\noalign{}
\endhead
\bottomrule\noalign{}
\endlastfoot
AAP & 66 ms, 27 MB & 13310 ms, 48585 MB \\
AME (all variables) & 8.9 s, 14.2 GB & Failed (``vector memory limit of
48.0 Gb reached'') \\
\end{longtable}

The 500,000-observation results demonstrate the fundamental scaling
difference: \textbf{{Margins.jl}} successfully completes AME computation
across all 65 variables while {marginaleffects} terminates with a memory
error before completing. The operations that \texttt{marginaleffects}
completed before failure showed memory ratios exceeding 1,791x, at 47.4
GB vs 27 MB for average adjusted predictions (see Appendix C for
hardware details).

\subsection{FormulaCompiler.jl
Micro-Benchmarks}\label{formulacompiler.jl-micro-benchmarks}

The performance characteristics above derive from
\textbf{{FormulaCompiler.jl}}'s zero-allocation evaluation. We verify
these guarantees through micro-benchmarks measuring per-row evaluation
timing and memory allocation (see Table~\ref{tbl-micro-benchmarks} for
formula specifications). Table~\ref{tbl-perrow-eval} demonstrates
constant-time evaluation across dataset sizes, and
Table~\ref{tbl-formula-complexity} shows scaling with formula
complexity.

Per-row evaluation (interaction-heavy formula):

\begin{longtable}[]{@{}llll@{}}
\caption{Per-row evaluation timing across dataset sizes, demonstrating
\(O(1)\) complexity in \(n\) with zero
allocations.}\label{tbl-perrow-eval}\tabularnewline
\toprule\noalign{}
\(n\) (rows) & Median (ns) & Min (ns) & Memory (bytes) \\
\midrule\noalign{}
\endfirsthead
\toprule\noalign{}
\(n\) (rows) & Median (ns) & Min (ns) & Memory (bytes) \\
\midrule\noalign{}
\endhead
\bottomrule\noalign{}
\endlastfoot
10,000 & 13.6 & 12.1 & 0 \\
100,000 & 13.6 & 12.1 & 0 \\
1,000,000 & 13.7 & 12.1 & 0 \\
\end{longtable}

Per-row evaluation time remains constant as dataset size increases by
two orders of magnitude, demonstrating \(O(1)\) complexity in \(n\).
Memory allocation is exactly zero bytes after warmup.

Formula complexity:

\begin{longtable}[]{@{}lll@{}}
\caption{Per-row evaluation timing by formula complexity, showing
\(O(p)\) scaling with zero
allocations.}\label{tbl-formula-complexity}\tabularnewline
\toprule\noalign{}
Formula & Median (ns) & Memory (bytes) \\
\midrule\noalign{}
\endfirsthead
\toprule\noalign{}
Formula & Median (ns) & Memory (bytes) \\
\midrule\noalign{}
\endhead
\bottomrule\noalign{}
\endlastfoot
Simple (3 terms) & 6.8 & 0 \\
Complex (interactions) & 13.6 & 0 \\
\end{longtable}

\begin{figure}

\centering{

\pandocbounded{\includegraphics[keepaspectratio]{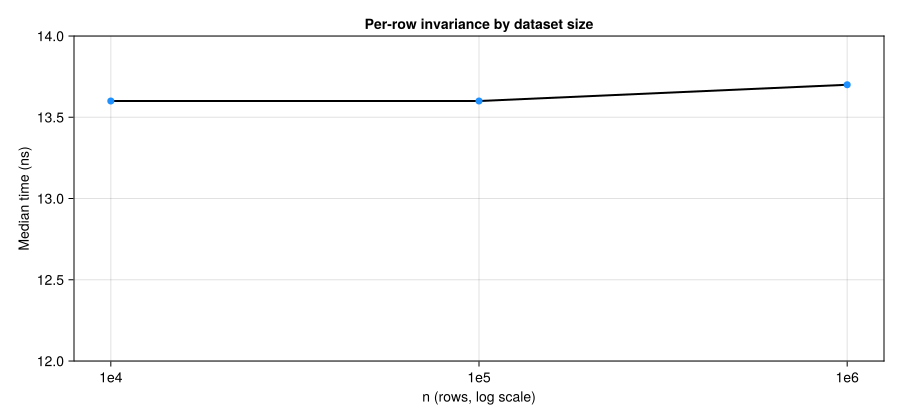}}

}

\caption{\label{fig-scaling}Per-row evaluation time remains constant as
\(n\) increases from 10,000 to 1,000,000 observations, demonstrating
\(O(1)\) complexity in dataset size.}

\end{figure}

Evaluation time scales with formula complexity (output width \(p\))
while maintaining zero allocations for all formula types.
Figure~\ref{fig-scaling} illustrates this constant-time behavior as
dataset size increases by two orders of magnitude.

\section{Case Studies}\label{case-studies}

We present four case studies demonstrating typical \textbf{{Margins.jl}}
workflows: scenario-based counterfactuals, interaction testing for
categorical and continuous moderators, and mixed model integration.

\subsection{Case Study 1: Policy
Counterfactuals}\label{case-study-1-policy-counterfactuals}

This case demonstrates counterfactual scenario analysis---computing how
effects change under hypothetical policy interventions.

Model: Linear model with treatment-covariate interaction:

\begin{Shaded}
\begin{Highlighting}[]
\ImportTok{using} \BuiltInTok{Margins}\NormalTok{, }\BuiltInTok{GLM}\NormalTok{, }\BuiltInTok{DataFrames}

\NormalTok{model }\OperatorTok{=} \FunctionTok{lm}\NormalTok{(}\PreprocessorTok{@formula}\NormalTok{(y }\OperatorTok{\textasciitilde{}}\NormalTok{ x }\OperatorTok{+}\NormalTok{ z }\OperatorTok{+}\NormalTok{ x }\OperatorTok{*}\NormalTok{ treatment), data)}

\CommentTok{\# Baseline: average marginal effect of x}
\NormalTok{baseline }\OperatorTok{=} \FunctionTok{population\_margins}\NormalTok{(model, data; }\KeywordTok{type}\OperatorTok{=:}\NormalTok{effects, vars}\OperatorTok{=}\NormalTok{[}\OperatorTok{:}\NormalTok{x])}

\CommentTok{\# Policy scenario: AME of x when treatment = 1 for everyone}
\NormalTok{policy }\OperatorTok{=} \FunctionTok{population\_margins}\NormalTok{(model, data; }\KeywordTok{type}\OperatorTok{=:}\NormalTok{effects, vars}\OperatorTok{=}\NormalTok{[}\OperatorTok{:}\NormalTok{x], scenarios}\OperatorTok{=}\NormalTok{(treatment}\OperatorTok{=}\FloatTok{1}\NormalTok{,))}
\end{Highlighting}
\end{Shaded}

The baseline AME of \texttt{x} under observed conditions:

\begin{Shaded}
\begin{Highlighting}[]
\NormalTok{EffectsResult}\OperatorTok{:} \FloatTok{1}\NormalTok{ population effects (N}\OperatorTok{=}\FloatTok{5000}\NormalTok{)}
\OperatorTok{{-}{-}{-}{-}{-}{-}{-}{-}{-}{-}{-}{-}{-}{-}{-}{-}{-}{-}{-}{-}{-}{-}{-}{-}{-}{-}{-}{-}{-}{-}{-}{-}{-}{-}{-}{-}{-}{-}{-}{-}{-}{-}{-}{-}{-}{-}{-}{-}{-}{-}{-}{-}{-}{-}{-}{-}{-}{-}{-}{-}{-}{-}{-}{-}{-}{-}{-}{-}{-}{-}{-}}
\NormalTok{Variable Contrast         dy}\OperatorTok{/}\NormalTok{dx   Std. Err.  [}\FloatTok{95}\OperatorTok{\%}\NormalTok{ Conf.   Interval]}
\OperatorTok{{-}{-}{-}{-}{-}{-}{-}{-}{-}{-}{-}{-}{-}{-}{-}{-}{-}{-}{-}{-}{-}{-}{-}{-}{-}{-}{-}{-}{-}{-}{-}{-}{-}{-}{-}{-}{-}{-}{-}{-}{-}{-}{-}{-}{-}{-}{-}{-}{-}{-}{-}{-}{-}{-}{-}{-}{-}{-}{-}{-}{-}{-}{-}{-}{-}{-}{-}{-}{-}{-}{-}}
\NormalTok{x        dy}\OperatorTok{/}\NormalTok{dx          }\FloatTok{0.36917}       \FloatTok{0.007}     \FloatTok{0.35544}      \FloatTok{0.3829}
\OperatorTok{{-}{-}{-}{-}{-}{-}{-}{-}{-}{-}{-}{-}{-}{-}{-}{-}{-}{-}{-}{-}{-}{-}{-}{-}{-}{-}{-}{-}{-}{-}{-}{-}{-}{-}{-}{-}{-}{-}{-}{-}{-}{-}{-}{-}{-}{-}{-}{-}{-}{-}{-}{-}{-}{-}{-}{-}{-}{-}{-}{-}{-}{-}{-}{-}{-}{-}{-}{-}{-}{-}{-}}
\end{Highlighting}
\end{Shaded}

Under the counterfactual policy scenario where all observations receive
treatment:

\begin{Shaded}
\begin{Highlighting}[]
\NormalTok{EffectsResult}\OperatorTok{:} \FloatTok{1}\NormalTok{ population effects (N}\OperatorTok{=}\FloatTok{5000}\NormalTok{)}
\NormalTok{Scenarios}\OperatorTok{:}\NormalTok{ treatment}
\OperatorTok{{-}{-}{-}{-}{-}{-}{-}{-}{-}{-}{-}{-}{-}{-}{-}{-}{-}{-}{-}{-}{-}{-}{-}{-}{-}{-}{-}{-}{-}{-}{-}{-}{-}{-}{-}{-}{-}{-}{-}{-}{-}{-}{-}{-}{-}{-}{-}{-}{-}{-}{-}{-}{-}{-}{-}{-}{-}{-}{-}{-}{-}{-}{-}{-}{-}{-}{-}{-}{-}{-}{-}{-}{-}{-}{-}{-}{-}{-}{-}{-}{-}{-}{-}{-}{-}{-}}
\NormalTok{Variable Contrast    at\_treatment        dy}\OperatorTok{/}\NormalTok{dx   Std. Err.  [}\FloatTok{95}\OperatorTok{\%}\NormalTok{ Conf.   Interval]}
\OperatorTok{{-}{-}{-}{-}{-}{-}{-}{-}{-}{-}{-}{-}{-}{-}{-}{-}{-}{-}{-}{-}{-}{-}{-}{-}{-}{-}{-}{-}{-}{-}{-}{-}{-}{-}{-}{-}{-}{-}{-}{-}{-}{-}{-}{-}{-}{-}{-}{-}{-}{-}{-}{-}{-}{-}{-}{-}{-}{-}{-}{-}{-}{-}{-}{-}{-}{-}{-}{-}{-}{-}{-}{-}{-}{-}{-}{-}{-}{-}{-}{-}{-}{-}{-}{-}{-}{-}}
\NormalTok{x        derivative  }\FloatTok{1}                 \FloatTok{0.43936}     \FloatTok{0.00999}     \FloatTok{0.41979}     \FloatTok{0.45894}
\OperatorTok{{-}{-}{-}{-}{-}{-}{-}{-}{-}{-}{-}{-}{-}{-}{-}{-}{-}{-}{-}{-}{-}{-}{-}{-}{-}{-}{-}{-}{-}{-}{-}{-}{-}{-}{-}{-}{-}{-}{-}{-}{-}{-}{-}{-}{-}{-}{-}{-}{-}{-}{-}{-}{-}{-}{-}{-}{-}{-}{-}{-}{-}{-}{-}{-}{-}{-}{-}{-}{-}{-}{-}{-}{-}{-}{-}{-}{-}{-}{-}{-}{-}{-}{-}{-}{-}{-}}
\end{Highlighting}
\end{Shaded}

Under observed conditions, the AME of \texttt{x} is 0.369. Under a
counterfactual policy where everyone receives treatment, the AME
increases to 0.439. The \texttt{scenarios} parameter enables
counterfactual analysis by fixing specific covariate values across all
observations while computing effects across the sample distribution.
This differs from the \texttt{groups} parameter, which computes separate
AME estimates for each level of a grouping variable within the observed
data. Use \texttt{scenarios} for ``what if'' questions (\emph{e.g.},
``What would the effect be if everyone received treatment?'') and
\texttt{groups} for subgroup analysis (\emph{e.g.}, ``How does the
effect differ by region?'').

\subsection{Case Study 2: GLM with Categorical
Interactions}\label{case-study-2-glm-with-categorical-interactions}

This case demonstrates marginal effects for a logistic regression with
continuous-categorical interactions, including formal testing via second
differences.

Model: Binary outcome with continuous predictor \texttt{x} interacting
with categorical \texttt{group}:

\begin{Shaded}
\begin{Highlighting}[]
\NormalTok{model }\OperatorTok{=} \FunctionTok{glm}\NormalTok{(}\PreprocessorTok{@formula}\NormalTok{(y }\OperatorTok{\textasciitilde{}}\NormalTok{ x }\OperatorTok{*}\NormalTok{ group }\OperatorTok{+}\NormalTok{ z), data, }\FunctionTok{Binomial}\NormalTok{())}

\CommentTok{\# 1. Main effects (AME across all variables)}
\NormalTok{ame\_result }\OperatorTok{=} \FunctionTok{population\_margins}\NormalTok{(model, data; }\KeywordTok{type}\OperatorTok{=:}\NormalTok{effects)}

\CommentTok{\# 2. Effect of x at each group level}
\NormalTok{ames\_by\_group }\OperatorTok{=} \FunctionTok{population\_margins}\NormalTok{(model, data;}
    \KeywordTok{type}\OperatorTok{=:}\NormalTok{effects, vars}\OperatorTok{=}\NormalTok{[}\OperatorTok{:}\NormalTok{x], scenarios}\OperatorTok{=}\NormalTok{(group}\OperatorTok{=}\NormalTok{[}\StringTok{"A"}\NormalTok{, }\StringTok{"B"}\NormalTok{, }\StringTok{"C"}\NormalTok{],))}

\CommentTok{\# 3. Test whether x effect differs by group (second differences)}
\NormalTok{sd }\OperatorTok{=} \FunctionTok{second\_differences}\NormalTok{(ames\_by\_group, }\OperatorTok{:}\NormalTok{x, }\OperatorTok{:}\NormalTok{group, }\FunctionTok{vcov}\NormalTok{(model))}
\end{Highlighting}
\end{Shaded}

The main effects (AME across all variables):

\begin{Shaded}
\begin{Highlighting}[]
\NormalTok{EffectsResult}\OperatorTok{:} \FloatTok{4}\NormalTok{ population effects (N}\OperatorTok{=}\FloatTok{5000}\NormalTok{)}
\OperatorTok{{-}{-}{-}{-}{-}{-}{-}{-}{-}{-}{-}{-}{-}{-}{-}{-}{-}{-}{-}{-}{-}{-}{-}{-}{-}{-}{-}{-}{-}{-}{-}{-}{-}{-}{-}{-}{-}{-}{-}{-}{-}{-}{-}{-}{-}{-}{-}{-}{-}{-}{-}{-}{-}{-}{-}{-}{-}{-}{-}{-}{-}{-}{-}{-}{-}{-}{-}{-}{-}{-}{-}}
\NormalTok{Variable Contrast         dy}\OperatorTok{/}\NormalTok{dx   Std. Err.  [}\FloatTok{95}\OperatorTok{\%}\NormalTok{ Conf.   Interval]}
\OperatorTok{{-}{-}{-}{-}{-}{-}{-}{-}{-}{-}{-}{-}{-}{-}{-}{-}{-}{-}{-}{-}{-}{-}{-}{-}{-}{-}{-}{-}{-}{-}{-}{-}{-}{-}{-}{-}{-}{-}{-}{-}{-}{-}{-}{-}{-}{-}{-}{-}{-}{-}{-}{-}{-}{-}{-}{-}{-}{-}{-}{-}{-}{-}{-}{-}{-}{-}{-}{-}{-}{-}{-}}
\NormalTok{x        dy}\OperatorTok{/}\NormalTok{dx          }\FloatTok{0.01457}     \FloatTok{0.00695}     \FloatTok{0.00096}     \FloatTok{0.02819}
\NormalTok{z        dy}\OperatorTok{/}\NormalTok{dx          }\FloatTok{0.01013}     \FloatTok{0.00694}    \OperatorTok{{-}}\FloatTok{0.00346}     \FloatTok{0.02372}
\NormalTok{group    B }\OperatorTok{{-}}\NormalTok{ A          }\FloatTok{0.02114}     \FloatTok{0.01702}    \OperatorTok{{-}}\FloatTok{0.01222}     \FloatTok{0.05451}
\NormalTok{group    C }\OperatorTok{{-}}\NormalTok{ A          }\FloatTok{0.07478}     \FloatTok{0.01697}     \FloatTok{0.04152}     \FloatTok{0.10805}
\OperatorTok{{-}{-}{-}{-}{-}{-}{-}{-}{-}{-}{-}{-}{-}{-}{-}{-}{-}{-}{-}{-}{-}{-}{-}{-}{-}{-}{-}{-}{-}{-}{-}{-}{-}{-}{-}{-}{-}{-}{-}{-}{-}{-}{-}{-}{-}{-}{-}{-}{-}{-}{-}{-}{-}{-}{-}{-}{-}{-}{-}{-}{-}{-}{-}{-}{-}{-}{-}{-}{-}{-}{-}}
\end{Highlighting}
\end{Shaded}

The effect of \texttt{x} varies by group level:

\begin{Shaded}
\begin{Highlighting}[]
\NormalTok{EffectsResult}\OperatorTok{:} \FloatTok{3}\NormalTok{ population effects (N}\OperatorTok{=}\FloatTok{5000}\NormalTok{)}
\NormalTok{Scenarios}\OperatorTok{:}\NormalTok{ group}
\OperatorTok{{-}{-}{-}{-}{-}{-}{-}{-}{-}{-}{-}{-}{-}{-}{-}{-}{-}{-}{-}{-}{-}{-}{-}{-}{-}{-}{-}{-}{-}{-}{-}{-}{-}{-}{-}{-}{-}{-}{-}{-}{-}{-}{-}{-}{-}{-}{-}{-}{-}{-}{-}{-}{-}{-}{-}{-}{-}{-}{-}{-}{-}{-}{-}{-}{-}{-}{-}{-}{-}{-}{-}{-}{-}{-}{-}{-}{-}{-}{-}{-}{-}{-}}
\NormalTok{Variable Contrast    at\_group        dy}\OperatorTok{/}\NormalTok{dx   Std. Err.  [}\FloatTok{95}\OperatorTok{\%}\NormalTok{ Conf.   Interval]}
\OperatorTok{{-}{-}{-}{-}{-}{-}{-}{-}{-}{-}{-}{-}{-}{-}{-}{-}{-}{-}{-}{-}{-}{-}{-}{-}{-}{-}{-}{-}{-}{-}{-}{-}{-}{-}{-}{-}{-}{-}{-}{-}{-}{-}{-}{-}{-}{-}{-}{-}{-}{-}{-}{-}{-}{-}{-}{-}{-}{-}{-}{-}{-}{-}{-}{-}{-}{-}{-}{-}{-}{-}{-}{-}{-}{-}{-}{-}{-}{-}{-}{-}{-}{-}}
\NormalTok{x        derivative  A             }\FloatTok{0.02404}     \FloatTok{0.01203}     \FloatTok{0.00045}     \FloatTok{0.04762}
\NormalTok{x        derivative  B             }\FloatTok{0.03503}     \FloatTok{0.01227}     \FloatTok{0.01098}     \FloatTok{0.05908}
\NormalTok{x        derivative  C            }\OperatorTok{{-}}\FloatTok{0.01419}      \FloatTok{0.0118}    \OperatorTok{{-}}\FloatTok{0.03732}     \FloatTok{0.00894}
\OperatorTok{{-}{-}{-}{-}{-}{-}{-}{-}{-}{-}{-}{-}{-}{-}{-}{-}{-}{-}{-}{-}{-}{-}{-}{-}{-}{-}{-}{-}{-}{-}{-}{-}{-}{-}{-}{-}{-}{-}{-}{-}{-}{-}{-}{-}{-}{-}{-}{-}{-}{-}{-}{-}{-}{-}{-}{-}{-}{-}{-}{-}{-}{-}{-}{-}{-}{-}{-}{-}{-}{-}{-}{-}{-}{-}{-}{-}{-}{-}{-}{-}{-}{-}}
\end{Highlighting}
\end{Shaded}

The second differences test whether these group-specific effects differ
significantly:

\begin{Shaded}
\begin{Highlighting}[]
\NormalTok{ Levels    }\FunctionTok{Delta}\NormalTok{(dy}\OperatorTok{/}\NormalTok{dx)     SE       z       p}\OperatorTok{{-}}\NormalTok{value}
\NormalTok{ A vs B    }\FloatTok{0.0110}    \FloatTok{0.0172}   \FloatTok{0.64}     \FloatTok{0.522}
\NormalTok{ A vs C    }\OperatorTok{{-}}\FloatTok{0.0382}    \FloatTok{0.0168}   \OperatorTok{{-}}\FloatTok{2.27}     \FloatTok{0.023}
\NormalTok{ B vs C    }\OperatorTok{{-}}\FloatTok{0.0492}    \FloatTok{0.0170}   \OperatorTok{{-}}\FloatTok{2.89}     \FloatTok{0.004}
\end{Highlighting}
\end{Shaded}

The marginal effect of \texttt{x} varies across group levels: positive
in groups A and B, negative in group C. The second differences test
whether these differences are statistically significant. Here, the A vs
C and B vs C contrasts reach significance at the 5\% level
(\(p < 0.05\)), indicating that the effect of \texttt{x} differs
significantly between group C and the other groups on the probability
scale.

\subsection{Case Study 3: Mixed Models Fixed
Effects}\label{case-study-3-mixed-models-fixed-effects}

This case demonstrates marginal effects for mixed-effects models,
focusing on fixed effects while accounting for the random effects
structure. The package integrates with {MixedModels.jl} and extracts the
fixed effects from the model to calculate the marginal effects.

Model: Linear mixed model with random intercepts:

\begin{Shaded}
\begin{Highlighting}[]
\ImportTok{using} \BuiltInTok{MixedModels}

\NormalTok{model }\OperatorTok{=} \FunctionTok{fit}\NormalTok{(MixedModel, }\PreprocessorTok{@formula}\NormalTok{(y }\OperatorTok{\textasciitilde{}}\NormalTok{ x }\OperatorTok{+}\NormalTok{ treatment }\OperatorTok{+}\NormalTok{ (}\FloatTok{1}\OperatorTok{|}\NormalTok{group)), data)}

\CommentTok{\# Fixed effects marginal effects}
\NormalTok{fe\_result }\OperatorTok{=} \FunctionTok{population\_margins}\NormalTok{(model, data; }\KeywordTok{type}\OperatorTok{=:}\NormalTok{effects)}
\FunctionTok{DataFrame}\NormalTok{(fe\_result)}
\end{Highlighting}
\end{Shaded}

\begin{Shaded}
\begin{Highlighting}[]
\NormalTok{EffectsResult}\OperatorTok{:} \FloatTok{2}\NormalTok{ population effects (N}\OperatorTok{=}\FloatTok{10000}\NormalTok{)}
\OperatorTok{{-}{-}{-}{-}{-}{-}{-}{-}{-}{-}{-}{-}{-}{-}{-}{-}{-}{-}{-}{-}{-}{-}{-}{-}{-}{-}{-}{-}{-}{-}{-}{-}{-}{-}{-}{-}{-}{-}{-}{-}{-}{-}{-}{-}{-}{-}{-}{-}{-}{-}{-}{-}{-}{-}{-}{-}{-}{-}{-}{-}{-}{-}{-}{-}{-}{-}{-}{-}{-}{-}{-}{-}{-}{-}{-}{-}}
\NormalTok{Variable   Contrast            dy}\OperatorTok{/}\NormalTok{dx   Std. Err.  [}\FloatTok{95}\OperatorTok{\%}\NormalTok{ Conf.   Interval]}
\OperatorTok{{-}{-}{-}{-}{-}{-}{-}{-}{-}{-}{-}{-}{-}{-}{-}{-}{-}{-}{-}{-}{-}{-}{-}{-}{-}{-}{-}{-}{-}{-}{-}{-}{-}{-}{-}{-}{-}{-}{-}{-}{-}{-}{-}{-}{-}{-}{-}{-}{-}{-}{-}{-}{-}{-}{-}{-}{-}{-}{-}{-}{-}{-}{-}{-}{-}{-}{-}{-}{-}{-}{-}{-}{-}{-}{-}{-}}
\NormalTok{x          dy}\OperatorTok{/}\NormalTok{dx             }\FloatTok{0.30348}      \FloatTok{0.0051}     \FloatTok{0.29348}     \FloatTok{0.31348}
\NormalTok{treatment  }\ConstantTok{true} \OperatorTok{{-}} \ConstantTok{false}      \FloatTok{0.49767}     \FloatTok{0.01011}     \FloatTok{0.47785}     \FloatTok{0.51749}
\OperatorTok{{-}{-}{-}{-}{-}{-}{-}{-}{-}{-}{-}{-}{-}{-}{-}{-}{-}{-}{-}{-}{-}{-}{-}{-}{-}{-}{-}{-}{-}{-}{-}{-}{-}{-}{-}{-}{-}{-}{-}{-}{-}{-}{-}{-}{-}{-}{-}{-}{-}{-}{-}{-}{-}{-}{-}{-}{-}{-}{-}{-}{-}{-}{-}{-}{-}{-}{-}{-}{-}{-}{-}{-}{-}{-}{-}{-}}
\end{Highlighting}
\end{Shaded}

Note that \textbf{{Margins.jl}} computes marginal effects for fixed
effects only, ignoring the random effects component.

\subsection{\texorpdfstring{Case Study 4: Continuous \(\times\)
Continuous
Interactions}{Case Study 4: Continuous \textbackslash times Continuous Interactions}}\label{case-study-4-continuous-times-continuous-interactions}

This case demonstrates the use of \texttt{second\_differences\_at} to
test whether the effect of one continuous variable changes with another
continuous variable. We show two scenarios: a linear interaction
(constant rate of change) and a quadratic interaction (varying rate of
change).

\textbf{Linear interaction:}

\begin{Shaded}
\begin{Highlighting}[]
\NormalTok{model }\OperatorTok{=} \FunctionTok{glm}\NormalTok{(}\PreprocessorTok{@formula}\NormalTok{(y }\OperatorTok{\textasciitilde{}}\NormalTok{ x }\OperatorTok{*}\NormalTok{ z), data, }\FunctionTok{Binomial}\NormalTok{())}

\NormalTok{sd }\OperatorTok{=} \FunctionTok{second\_differences\_at}\NormalTok{(model, data, }\OperatorTok{:}\NormalTok{x, }\OperatorTok{:}\NormalTok{z, }\FunctionTok{vcov}\NormalTok{(model);}
\NormalTok{    at}\OperatorTok{=}\FunctionTok{quantile}\NormalTok{(data.z, [}\FloatTok{0.25}\NormalTok{, }\FloatTok{0.5}\NormalTok{, }\FloatTok{0.75}\NormalTok{]))}
\end{Highlighting}
\end{Shaded}

\begin{Shaded}
\begin{Highlighting}[]
\NormalTok{ eval\_point (z)   }\FunctionTok{d}\NormalTok{(AME\_x)}\OperatorTok{/}\NormalTok{dz      SE        z       p}\OperatorTok{{-}}\NormalTok{value}
 \FloatTok{0.41}\NormalTok{ (}\FloatTok{25}\NormalTok{th }\OperatorTok{\%}\NormalTok{ile)   }\FloatTok{0.0162}      \FloatTok{0.0111}    \FloatTok{1.46}      \FloatTok{0.144}
 \FloatTok{0.77}\NormalTok{ (}\FloatTok{50}\NormalTok{th }\OperatorTok{\%}\NormalTok{ile)   }\FloatTok{0.0155}      \FloatTok{0.0113}    \FloatTok{1.37}      \FloatTok{0.169}
 \FloatTok{1.26}\NormalTok{ (}\FloatTok{75}\NormalTok{th }\OperatorTok{\%}\NormalTok{ile)   }\FloatTok{0.0142}      \FloatTok{0.0113}    \FloatTok{1.26}      \FloatTok{0.209}
\end{Highlighting}
\end{Shaded}

With a linear interaction, the derivative
\(\partial(\text{AME})/\partial z\) is approximately constant across all
evaluation points---the effect of \texttt{x} changes at a similar rate
regardless of where we evaluate along \texttt{z}.

\textbf{Quadratic interaction:}

\begin{Shaded}
\begin{Highlighting}[]
\NormalTok{model }\OperatorTok{=} \FunctionTok{glm}\NormalTok{(}\PreprocessorTok{@formula}\NormalTok{(y }\OperatorTok{\textasciitilde{}}\NormalTok{ x }\OperatorTok{*}\NormalTok{ z }\OperatorTok{+}\NormalTok{ x }\OperatorTok{*}\NormalTok{ z2), data, }\FunctionTok{Binomial}\NormalTok{())}

\NormalTok{sd }\OperatorTok{=} \FunctionTok{second\_differences\_at}\NormalTok{(model, data, }\OperatorTok{:}\NormalTok{x, }\OperatorTok{:}\NormalTok{z, }\FunctionTok{vcov}\NormalTok{(model);}
\NormalTok{    at}\OperatorTok{=}\NormalTok{[}\FloatTok{0.5}\NormalTok{, }\FloatTok{1.5}\NormalTok{, }\FloatTok{2.5}\NormalTok{])}
\end{Highlighting}
\end{Shaded}

\begin{Shaded}
\begin{Highlighting}[]
\NormalTok{ eval\_point (z)   }\FunctionTok{d}\NormalTok{(AME\_x)}\OperatorTok{/}\NormalTok{dz      SE        z       p}\OperatorTok{{-}}\NormalTok{value}
 \FloatTok{0.5}                \FloatTok{0.0146}      \FloatTok{0.0365}    \FloatTok{0.40}      \FloatTok{0.689}
 \FloatTok{1.5}                \FloatTok{0.0141}      \FloatTok{0.0349}    \FloatTok{0.40}      \FloatTok{0.687}
 \FloatTok{2.5}                \FloatTok{0.0133}      \FloatTok{0.0326}    \FloatTok{0.41}      \FloatTok{0.682}
\end{Highlighting}
\end{Shaded}

With a quadratic interaction, the derivative varies across evaluation
points: the rate at which \texttt{x}'s effect changes itself depends on
the value of \texttt{z}. The \texttt{second\_differences\_at} function
detects this curvature by evaluating the local derivative at multiple
points along the moderator's range.

\section{Future Work}\label{future-work}

Several methodological directions could extend this work. First, while
\textbf{{Margins.jl}} currently computes marginal effects for fixed
effects only, a more comprehensive treatment would propagate random
effect uncertainty into the marginal effects themselves. This would
require integrating over the random effect distribution, either
analytically for simple designs or through simulation for more complex
hierarchical structures (Skrondal and Rabe-Hesketh 2009). This would
enable inference for marginal effects in clustered and longitudinal
designs where random effects capture substantive heterogeneity.

Additionally, the package could extend support to additional model
classes. Survival models require computing marginal effects on the
survival probability scale, instead of the hazard ratio scale, which
involves integration over the censoring distribution. Ordinal regression
models yield category-specific marginal effects---the change in
probability for each outcome level---requiring careful interpretation
when categories have natural ordering. Multinomial choice models involve
cross-alternative effects where changing one covariate affects
probabilities across all alternatives, with implications for the
independence of irrelevant alternatives assumption.

While \textbf{{Margins.jl}} supports second differences for two-way
interactions, the package could also be extended to directly support the
analysis of higher-order interactions. The connection to treatment
effect heterogeneity may be relevant: marginal effects that vary across
the covariate distribution can identify subgroups for whom interventions
differ in effectiveness, linking to the causal inference literature on
conditional average treatment effects. Finally, the package could be
extended to support parallel computation to handle even more complex
models with larger datasets, particularly for simulation-based random
effects integration and heterogeneity analysis with large datasets.

\section{Validation}\label{validation}

Both packages undergo comprehensive automated testing to ensure
numerical correctness and performance guarantees. \textbf{{Margins.jl}}
validates marginal effects computations against R's {marginaleffects}
package using identical datasets exported across languages, with test
tolerances requiring agreement within 0.01\% for point estimates and
0.1\% for standard errors. The test suite covers all combinations of the
\(2 \times 2\) framework, categorical contrasts, interaction effects,
and elasticity measures. \textbf{{FormulaCompiler.jl}} validates
compiled formula output against {StatsModels.jl}'s
\texttt{modelmatrix()} reference implementation, and zero-allocation
guarantees are verified through BenchmarkTools (Chen and Revels 2024)
measurements. Both packages use continuous integration with automated
test execution on each commit.

\section{Conclusion}\label{conclusion}

This paper presents \textbf{{Margins.jl}} and
\textbf{{FormulaCompiler.jl}}, two {Julia} packages for marginal effects
computation. \textbf{{Margins.jl}} provides a statistical interface
organized around a \(2 \times 2\) framework distinguishing population
from profile analysis and effects from predictions.
\textbf{{FormulaCompiler.jl}} provides the computational foundation
through position-mapped formula compilation that achieves
zero-allocation evaluation.

The architectural separation between statistical interface and
computational engine allows each layer to be optimized independently.
Users interact with familiar statistical concepts while the underlying
implementation handles performance considerations through
type-specialized compilation and pre-allocated buffers.

Benchmarks show substantial performance improvements over existing
implementations: 622x average speedup relative to R's {marginaleffects}
package, 460x memory reduction, and successful computation at dataset
sizes where comparison implementations exceed available memory. These
characteristics make average marginal effects computation practical at
scales where marginal effects at the mean or omitting standard errors
might otherwise be necessary compromises.

The computational techniques in \textbf{{FormulaCompiler.jl}}
(type-specialized compilation, pre-allocated buffers, orchestrated
automatic differentiation) apply beyond marginal effects to other
workloads requiring repeated formula evaluation. Monte Carlo simulation,
bootstrap inference, cross-validation, sensitivity analysis, and power
analysis share this computational pattern, where per-evaluation overhead
compounds across many iterations.

\section{Software Availability}\label{software-availability}

Both packages are registered in Julia's General Registry and can be
installed with \texttt{Pkg.add()}. Additionally, the development
versions may be installed directly from GitHub.

\textbf{{Margins.jl}}:

\begin{itemize}
\tightlist
\item
  Repository: \url{https://github.com/emfeltham/Margins.jl}
\item
  Installation: \texttt{using\ Pkg;\ Pkg.add("Margins")}
\item
  Documentation: \url{https://emfeltham.github.io/Margins.jl/}
\end{itemize}

\textbf{{FormulaCompiler.jl}}:

\begin{itemize}
\tightlist
\item
  Repository: \url{https://github.com/emfeltham/FormulaCompiler.jl}
\item
  Installation: \texttt{using\ Pkg;\ Pkg.add("FormulaCompiler")}
\item
  Documentation: \url{https://emfeltham.github.io/FormulaCompiler.jl/}
\end{itemize}

Both packages are MIT licensed and integrate with {Julia}'s statistical
ecosystem ({GLM.jl}, {MixedModels.jl}, {StatsModels.jl},
{DataFrames.jl}).

\section{Appendix A: Finite Difference Method for Continuous
Interactions}\label{appendix-a-finite-difference-method-for-continuous-interactions}

The \texttt{second\_differences\_at()} function computes local
derivatives of marginal effects with respect to continuous moderators
using symmetric finite differences, rather than automatic
differentiation. For a marginal effect \(\text{AME}(z)\) evaluated at
moderator value \(z\):

\[\frac{\partial \text{AME}}{\partial z} \approx \frac{\text{AME}(z + \delta) - \text{AME}(z - \delta)}{2\delta}\]

Step size selection: By default, \(\delta = 0.01 \times \text{SD}(z)\),
where \(\text{SD}(z)\) is the sample standard deviation of the
moderator. This balances two sources of error: (1) truncation error
\(O(\delta^2)\) from the finite difference approximation, and
floating-point error from catastrophic cancellation when \(\delta\) is
too small.

Numerical stability: For evaluation points with large magnitude
(\(|z| > 1\)), the step size is bounded below by \(|z| \times 10^{-8}\)
to ensure numerical distinctness of \(z \pm \delta\). For small values
near zero, a minimum absolute \(\delta\) of \(10^{-9}\) prevents
degenerate cases.

Standard errors: The delta method propagates parameter uncertainty
through the finite difference:

\[\text{SE}\left(\frac{\partial \text{AME}}{\partial z}\right) = \frac{\sqrt{(g_+ - g_-)^T \Sigma (g_+ - g_-)}}{2\delta}\]

where \(g_+\) and \(g_-\) are the gradient vectors of
\(\text{AME}(z + \delta)\) and \(\text{AME}(z - \delta)\) respectively,
and \(\Sigma\) is the parameter covariance matrix.

\section{Appendix B: Benchmark
Specifications}\label{appendix-b-benchmark-specifications}

The benchmark formulas and dataset sizes represent complex applied
analyses that require computing marginal effects for complex GLM/GLMM
specifications (interactions, nonlinear transforms, categorical
contrasts) on datasets with \(n > 50,000\) observations.

\subsection{FormulaCompiler
Micro-Benchmarks}\label{formulacompiler-micro-benchmarks}

These benchmarks use synthetic data with 5 variables: \texttt{y}
(continuous), \texttt{x} (continuous), \texttt{z} (positive continuous),
\texttt{group} (categorical A/B/C), \texttt{b} (boolean).

Source: \texttt{scripts/benchmarks.jl}.

\begin{longtable}[]{@{}
  >{\raggedright\arraybackslash}p{(\linewidth - 6\tabcolsep) * \real{0.3667}}
  >{\raggedright\arraybackslash}p{(\linewidth - 6\tabcolsep) * \real{0.1000}}
  >{\raggedright\arraybackslash}p{(\linewidth - 6\tabcolsep) * \real{0.3000}}
  >{\raggedright\arraybackslash}p{(\linewidth - 6\tabcolsep) * \real{0.2333}}@{}}
\caption{FormulaCompiler micro-benchmark
specifications.}\label{tbl-micro-benchmarks}\tabularnewline
\toprule\noalign{}
\begin{minipage}[b]{\linewidth}\raggedright
Benchmark
\end{minipage} & \begin{minipage}[b]{\linewidth}\raggedright
\(n\)
\end{minipage} & \begin{minipage}[b]{\linewidth}\raggedright
Formula
\end{minipage} & \begin{minipage}[b]{\linewidth}\raggedright
Model
\end{minipage} \\
\midrule\noalign{}
\endfirsthead
\toprule\noalign{}
\begin{minipage}[b]{\linewidth}\raggedright
Benchmark
\end{minipage} & \begin{minipage}[b]{\linewidth}\raggedright
\(n\)
\end{minipage} & \begin{minipage}[b]{\linewidth}\raggedright
Formula
\end{minipage} & \begin{minipage}[b]{\linewidth}\raggedright
Model
\end{minipage} \\
\midrule\noalign{}
\endhead
\bottomrule\noalign{}
\endlastfoot
Core row evaluation & 50,000 &
\texttt{y\ \textasciitilde{}\ x\ *\ group\ +\ log1p(z)} & OLS \\
Allocation comparison & 10,000 &
\texttt{y\ \textasciitilde{}\ x\ *\ group\ +\ log1p(z)} & OLS \\
Scenario overhead & 50,000 &
\texttt{y\ \textasciitilde{}\ x\ *\ group\ +\ log1p(z)} & OLS \\
Derivatives & 20,000 &
\texttt{y\ \textasciitilde{}\ x\ *\ group\ +\ log1p(z)} & OLS \\
Marginal effects (\(\eta\)/\(\mu\)) & 20,000 &
\texttt{y\ \textasciitilde{}\ x\ *\ group\ +\ log1p(z)} & OLS \\
Delta method SE & 5,000 &
\texttt{y\ \textasciitilde{}\ x\ *\ group\ +\ log1p(z)} & OLS \\
MixedModels & 10,000 &
\texttt{y\ \textasciitilde{}\ x\ +\ (1\ \textbar{}\ group)} & LMM \\
Simple formula & 30,000 &
\texttt{y\ \textasciitilde{}\ x\ +\ z\ +\ group} & OLS \\
Complex formula & 30,000 &
\texttt{y\ \textasciitilde{}\ x\ *\ group\ +\ log1p(z)\ +\ x\ \&\ z\ +\ x\ \&\ log1p(z)}
& OLS \\
Scale invariance (\(n\)) & 10K--1M &
\texttt{y\ \textasciitilde{}\ x\ *\ group\ +\ log1p(z)\ +\ x\ \&\ z\ +\ x\ \&\ log1p(z)}
& OLS \\
Mixtures & 20,000 &
\texttt{y\ \textasciitilde{}\ x\ *\ group\ +\ log1p(z)} & OLS \\
Population AME & 50,000 &
\texttt{y\ \textasciitilde{}\ x\ *\ group\ +\ log1p(z)} & OLS \\
Profile margins & 10K / 1M &
\texttt{y\ \textasciitilde{}\ x\ *\ group\ +\ log1p(z)} & OLS \\
Categorical AME (group) & 20,000 &
\texttt{y\ \textasciitilde{}\ x\ +\ group} & OLS \\
Categorical AME (bool) & 20,000 & \texttt{y\ \textasciitilde{}\ x\ +\ b}
& OLS \\
Parameter scaling & 10,000 &
\texttt{y\ \textasciitilde{}\ x1\ +\ ...\ +\ xp} (\(p=\{5,10,20,50\}\))
& OLS \\
\end{longtable}

\subsection{Case Studies}\label{case-studies-1}

\begin{longtable}[]{@{}
  >{\raggedright\arraybackslash}p{(\linewidth - 6\tabcolsep) * \real{0.2400}}
  >{\raggedright\arraybackslash}p{(\linewidth - 6\tabcolsep) * \real{0.1200}}
  >{\raggedright\arraybackslash}p{(\linewidth - 6\tabcolsep) * \real{0.3600}}
  >{\raggedright\arraybackslash}p{(\linewidth - 6\tabcolsep) * \real{0.2800}}@{}}
\caption{Case study
specifications.}\label{tbl-case-studies}\tabularnewline
\toprule\noalign{}
\begin{minipage}[b]{\linewidth}\raggedright
Case
\end{minipage} & \begin{minipage}[b]{\linewidth}\raggedright
\(n\)
\end{minipage} & \begin{minipage}[b]{\linewidth}\raggedright
Formula
\end{minipage} & \begin{minipage}[b]{\linewidth}\raggedright
Model
\end{minipage} \\
\midrule\noalign{}
\endfirsthead
\toprule\noalign{}
\begin{minipage}[b]{\linewidth}\raggedright
Case
\end{minipage} & \begin{minipage}[b]{\linewidth}\raggedright
\(n\)
\end{minipage} & \begin{minipage}[b]{\linewidth}\raggedright
Formula
\end{minipage} & \begin{minipage}[b]{\linewidth}\raggedright
Model
\end{minipage} \\
\midrule\noalign{}
\endhead
\bottomrule\noalign{}
\endlastfoot
GLM Logit + Interactions & 20,000 &
\texttt{y\ \textasciitilde{}\ x\ *\ group\ +\ log1p(z)} & Logistic \\
Linear Model + Scenarios & 50,000 &
\texttt{y\ \textasciitilde{}\ x\ +\ z\ +\ x\ *\ group} & OLS \\
MixedModels Fixed Effects & 10,000 &
\texttt{y\ \textasciitilde{}\ x\ +\ treatment\ +\ (1\ \textbar{}\ group)}
& LMM \\
\end{longtable}

\subsection{R Comparison Benchmark}\label{r-comparison-benchmark}

The performance comparison against R's marginaleffects package uses a
complex logistic regression model reflecting real-world analyses with
large synthetic data, modeled after Feltham, Forastiere, and Christakis
(2025):

\textbf{Dataset}: 5,000 observations (standard benchmark) / 500,000
observations (large benchmark), with \textasciitilde40 variables
(Table~\ref{tbl-bench-vars}).

\begin{longtable}[]{@{}
  >{\raggedright\arraybackslash}p{(\linewidth - 2\tabcolsep) * \real{0.4762}}
  >{\raggedright\arraybackslash}p{(\linewidth - 2\tabcolsep) * \real{0.5238}}@{}}
\caption{Variable categories in the R comparison benchmark
dataset.}\label{tbl-bench-vars}\tabularnewline
\toprule\noalign{}
\begin{minipage}[b]{\linewidth}\raggedright
Category
\end{minipage} & \begin{minipage}[b]{\linewidth}\raggedright
Variables
\end{minipage} \\
\midrule\noalign{}
\endfirsthead
\toprule\noalign{}
\begin{minipage}[b]{\linewidth}\raggedright
Category
\end{minipage} & \begin{minipage}[b]{\linewidth}\raggedright
Variables
\end{minipage} \\
\midrule\noalign{}
\endhead
\bottomrule\noalign{}
\endlastfoot
Network structure & \texttt{dists\_a\_inv}, \texttt{dists\_p\_inv},
\texttt{are\_related\_dists\_a\_inv} \\
Demographics & \texttt{age\_p}, \texttt{wealth\_d1\_4\_p},
\texttt{schoolyears\_p}, \texttt{man\_p} \\
Homophily measures & \texttt{hhi\_religion}, \texttt{hhi\_indigenous},
\texttt{age\_h}, \texttt{wealth\_d1\_4\_h} \\
Village characteristics & \texttt{population},
\texttt{coffee\_cultivation}, \texttt{market} \\
Categorical & \texttt{socio4}, \texttt{relation},
\texttt{religion\_c\_p} \\
\end{longtable}

\textbf{Model}: 65-parameter logistic regression (\texttt{Bernoulli},
\texttt{LogitLink}) with nested interactions:

\begin{Shaded}
\begin{Highlighting}[]
\PreprocessorTok{@formula}\NormalTok{(response }\OperatorTok{\textasciitilde{}}
\NormalTok{    socio4 }\OperatorTok{+}\NormalTok{ dists\_p\_inv }\OperatorTok{+}\NormalTok{ dists\_a\_inv }\OperatorTok{+}\NormalTok{ are\_related\_dists\_a\_inv }\OperatorTok{+}
\NormalTok{    socio4 }\OperatorTok{\&}\NormalTok{ (dists\_p\_inv }\OperatorTok{+}\NormalTok{ are\_related\_dists\_a\_inv }\OperatorTok{+}\NormalTok{ dists\_a\_inv) }\OperatorTok{+}
\NormalTok{    (age\_p }\OperatorTok{+}\NormalTok{ wealth\_d1\_4\_p }\OperatorTok{+}\NormalTok{ schoolyears\_p }\OperatorTok{+}\NormalTok{ man\_p }\OperatorTok{+}\NormalTok{ same\_building }\OperatorTok{+}
\NormalTok{     population }\OperatorTok{+}\NormalTok{ hhi\_religion }\OperatorTok{+}\NormalTok{ hhi\_indigenous }\OperatorTok{+}\NormalTok{ coffee\_cultivation }\OperatorTok{+}
\NormalTok{     market) }\OperatorTok{\&}\NormalTok{ (}\FloatTok{1} \OperatorTok{+}\NormalTok{ socio4 }\OperatorTok{+}\NormalTok{ are\_related\_dists\_a\_inv) }\OperatorTok{+}
\NormalTok{    relation }\OperatorTok{+}\NormalTok{ religion\_c\_p }\OperatorTok{+}\NormalTok{ relation }\OperatorTok{\&}\NormalTok{ socio4 }\OperatorTok{+}
\NormalTok{    religion\_c\_p }\OperatorTok{\&}\NormalTok{ are\_related\_dists\_a\_inv }\OperatorTok{+}
\NormalTok{    degree\_a\_mean }\OperatorTok{+}\NormalTok{ degree\_h }\OperatorTok{+}\NormalTok{ age\_a\_mean }\OperatorTok{+}\NormalTok{ schoolyears\_a\_mean }\OperatorTok{+}\NormalTok{ wealth\_d1\_4\_a\_mean }\OperatorTok{+}
\NormalTok{    age\_h }\OperatorTok{\&}\NormalTok{ age\_h\_nb\_1\_socio }\OperatorTok{+}
\NormalTok{    schoolyears\_h }\OperatorTok{\&}\NormalTok{ schoolyears\_h\_nb\_1\_socio }\OperatorTok{+}
\NormalTok{    wealth\_d1\_4\_h }\OperatorTok{\&}\NormalTok{ wealth\_d1\_4\_h\_nb\_1\_socio }\OperatorTok{+}
\NormalTok{    (degree\_a\_mean }\OperatorTok{+}\NormalTok{ degree\_h }\OperatorTok{+}\NormalTok{ age\_a\_mean }\OperatorTok{+}\NormalTok{ schoolyears\_a\_mean }\OperatorTok{+}
\NormalTok{     wealth\_d1\_4\_a\_mean) }\OperatorTok{\&}\NormalTok{ socio4 }\OperatorTok{+}
\NormalTok{    hhi\_religion }\OperatorTok{\&}\NormalTok{ are\_related\_dists\_a\_inv }\OperatorTok{+}
\NormalTok{    hhi\_indigenous }\OperatorTok{\&}\NormalTok{ are\_related\_dists\_a\_inv)}
\end{Highlighting}
\end{Shaded}

This specification includes binary \(\times\) continuous interactions,
individual-level variables with three-way interactions, categorical
interactions, and tie-level homophily measures.

Source: \texttt{Margins.jl/test/r\_compare/performance\_benchmark.jl}.

\section{Appendix C: Benchmark
Environment}\label{appendix-c-benchmark-environment}

All benchmarks were performed on the following system:

\begin{itemize}
\tightlist
\item
  Hardware: Apple M4 Pro (14 cores: 10 performance, 4 efficiency), 48 GB
  RAM
\item
  Operating system: macOS 15.6.1
\item
  Julia: 1.12.0
\item
  R: 4.5.1
\end{itemize}

Julia timing: {BenchmarkTools.jl} (Chen and Revels 2016) with sufficient
warmup iterations to ensure Just-In-Time (JIT) compilation; reported
values are minimum time across samples. Memory allocations are also
measured using {BenchmarkTools.jl} (Chen and Revels 2016).

R timing: The \texttt{microbenchmark} package (Mersmann et al. 2024)
with 5 iterations per operation (3 for large dataset); reported values
are median time. Memory measured using the \texttt{profmem} package
tracking total bytes allocated per operation.

Both benchmarks use identical datasets exported to ensure comparability.
All marginal effects are computed on the response scale with
delta-method standard errors. Table~\ref{tbl-micro-benchmarks} and
Table~\ref{tbl-case-studies} detail the benchmark formulas and dataset
sizes; complete benchmark scripts are available in the reproducibility
materials.

\section*{References}\label{references}
\addcontentsline{toc}{section}{References}

\phantomsection\label{refs}

\end{document}